\documentclass[aps,prl,reprint,groupedaddress,twocolumn]{revtex4-1}

\usepackage{xcolor}
\usepackage[utf8]{inputenc}
\usepackage{graphicx}
\usepackage{amsmath}
\usepackage{amssymb}
\usepackage{float}

\newcommand{\ket}[1]{\big|#1\big>}

\newcommand{\bk}{\mathbf{k}}

\newcommand{\bq}{\mathbf{q}}

\def\dz2{d$_{\text{z}^2}$}

\def\dx2y2{d$_{\text{x}^2\text{y}^2}$}
\def\G0W0{G$_0$W$_0$}
\def\scGW0{scGW$_0$}

\newcommand{\epstd}{\tilde\varepsilon}

\begin{document}



\title{Dynamical screening effects of substrate phonons on two-dimensional excitons}

%
%

 \author{Alexander Steinhoff}
 \affiliation{Institut f\"ur Theoretische Physik, Universit\"at Bremen, P.O. Box 330 440, 28334 Bremen, Germany}

 \author{Matthias Florian}
 \affiliation{Institut f\"ur Theoretische Physik, Universit\"at Bremen, P.O. Box 330 440, 28334 Bremen, Germany}

%

 \author{Frank Jahnke}
 \affiliation{Institut f\"ur Theoretische Physik, Universit\"at Bremen, P.O. Box 330 440, 28334 Bremen, Germany}
 \affiliation{MAPEX Center for Materials and Processes, Universit\"at Bremen, 28359 Bremen, Germany}





\begin{abstract}

Atomically thin materials are exceedingly susceptible to their dielectric environment. 
For transition metal dichalcogenides, sample placement on a substrate or encapsulation in hexagonal boron nitride (hBN) are frequently used.
In this paper we show that the
dielectric response due to optical phonons of adjacent materials influences 
excitons 
in
2d
crystals.
We provide an analytic model for the coupling of 2d charge carriers to optical substrate phonons, which causes polaron effects similar to that of intrinsic 2d phonons. 
We apply the model to 
hBN-encapsulated WSe$_2$,  
finding a significant reduction of the exciton binding energies 
due to dynamical screening effects.



\end{abstract}

\maketitle


Monolayers of transition metal dichalcogenide (TMD) semiconductors exhibit strong Coulomb interaction of their charge carriers giving rise to bound electron-hole states, known as excitons, with remarkable oscillator strength in optical spectra. \cite{qiu_optical_2013,chernikov_exciton_2014,steinhoff_influence_2014} Along with reciprocal-space valleys as a new optically addressable degree of freedom
\cite{xu_spin_2014}, this recommends TMD semiconductors as active materials in future optoelectronic devices such as light-emitting diodes \cite{pospischil_solar-energy_2014, baugher_optoelectronic_2014, ross_electrically_2014, withers_light-emitting_2015}, solar cells \cite{pospischil_solar-energy_2014, baugher_optoelectronic_2014}, and lasers \cite{wu_monolayer_2015, ye_monolayer_2015, salehzadeh_optically_2015, li_room-temperature_2017}. Key to these applications is the compatibility with different substrates or other two-dimensional (2d) materials in functional van der Waals heterostructures (vdW-HS). \cite{geim_van_2013} 
Fascinating prospects arise from the possibility to engineer electronic and optical properties by manipulation of the Coulomb interaction in atomically thin materials via its dielectric environment 
\cite{berkelbach_theory_2013, latini_excitons_2015, steinke_non-invasive_2017, trolle_model_2017, raja_coulomb_2017, steinhoff_exciton_2017,meckbach_influence_2018,florian_dielectric_2018,meckbach_giant_2018}.
It has also become customary to 
improve TMD sample qualities by means of encapsulation in hexagonal boron nitride (hBN). \cite{cadiz_excitonic_2017} 



Environmental screening is frequently described with a macroscopic model dielectric function of the vdW-HS formed by the TMD layer and adjacent layers. The so-called Rytova-Keldysh potential is a simple yet efficient workhorse, \cite{keldysh_coulomb_1979, rytova_screened_2018} where the environment is usually characterised by a static dielectric constant. It has been discussed recently that anomalous exciton binding energies in gallium oxide can be understood by means of dynamical screening from optical phonons going beyond a static Wannier picture. \cite{bechstedt_influence_2019}
Hence typical substrate materials such as sapphire, SiO$_2$ and hBN hosting optical phonons in the infrared spectral range are expected to add a significant frequency-dependent dielectric response felt by the encapsulated material. 
So far, the coupling of graphene plasmons to surface-optical phonon modes in a substrate has been considered. \cite{karimi_dielectric_2016, hwang_plasmon-phonon_2010} 
Substrate plasmons have been suggested as a tuning knob for the electronic properties of atomically thin layers. \cite{steinhoff_frequency-dependent_2018} Only recently substrate phonons have been discovered as an additional degree of freedom to tailor the electronic and optical properties of active TMD materials. \cite{chow_unusual_2017,jin_interlayer_2017}



In this paper, we provide a description 
for the coupling of TMD charge carriers to substrate phonons. Our approach 
uses a mapping of
the macroscopic dielectric function of the vdW-HS to an effective Fröhlich Hamiltonian. The Hamiltonian captures the microscopic parameters that characterize both, the heterostructure geometry and the properties of substrate TO phonons as extracted from experiments. Thereby, we transfer the dynamical, frequency-dependent behavior of the substrate dielectric function to a carrier-boson interaction on the level of second quantization. This puts the coupling to substrate phonons on an equal footing with the coupling to intrisic 2d phonons~\cite{selig_excitonic_2016} and introduces additional scattering channels for the 2d excitons. We then use the augmented coupling Hamiltonian in an equation-of-motion (EOM-) approach to investigate the impact on 2d excitons. In general, renormalizations of exciton binding energies and quasi-particle band gaps scale as an inverse power law with respect to the substrate TO phonon energy. For monolayer WSe$_2$ encapsulated in hBN, taking into account the anisotropy of the hBN dielectric response~\cite{geick_normal_1966}, we find that the coupling to hBN phonons leads to a significant reduction of the 1s-exciton binding energy in WSe$_2$. The effect becomes weaker for increasing exciton principal quantum number and is accompanied with additional line broadening.

To develop our theory of carriers coupling to substrate phonons, we start from the frequency-dependence of the substrate dielectric function. 
Specifically,
a Lorentz-oscillator model \cite{haug_electron_1984, karimi_dielectric_2016} is used, where the parameters can be either adjusted to fit experimental data or calculated from first principles:
\begin{equation}
 \begin{split}
 \varepsilon^{\textrm{s}}(\omega)=\varepsilon_{\infty}+\frac{s^2}{\omega_0^2-\omega^2-i\gamma\omega}\,.
 \end{split}
\label{eq:substrate_DF_TO}
\end{equation}
Here $\omega_0$ corresponds to the TO-phonon frequency, $s$ is the oscillator strength of the phonon and $\varepsilon_{\infty}$ is the high-frequency dielectric constant that takes into account screening due to inner-shell electrons in the substrate. While the high-frequency constant is commonly used to describe static screening in terms of a Rytova-Keldysh model, coupling to optical phonons introduces a dynamical screening contribution. 
In fact, carrier-phonon interaction is formally equivalent to a retarded screened Coulomb interaction from a diagrammatic point of view. \cite{leeuwen_first-principles_2004, bechstedt_influence_2019} Assuming infinitesimal damping of the phonons, we can identify the phonon propagator, see Eqs.~(116)-(117) in Ref.~\citenum{leeuwen_first-principles_2004} and Eq.(5) in Ref.~\citenum{bechstedt_influence_2019}, with the 
screened Coulomb interaction $W^{\textrm{HS}}_{\bq}(\omega)$:
%
\begin{equation}
 \begin{split}
 &\textrm{Im}\,W^{\textrm{HS}}_{\bq}(\omega)=V_{\bq}\textrm{Im}\,\varepsilon^{\textrm{HS},-1}_{\bq}(\omega) \\ &=-\pi \left|g_{\bq}\right|^2\left(\delta(\hbar\omega-\hbar\Omega_{\bq}) - \delta(\hbar\omega+\hbar\Omega_{\bq})\right)\,.
 \end{split}
\label{eq:propagator}
\end{equation}
Here $\varepsilon^{\textrm{HS}}_{\bq}(\omega)$ is the dielectric function for carriers in a 2d layer embedded in a heterostructure. Thus the loss function
$\textrm{Im}\,\varepsilon^{\textrm{HS},-1}_{\bq}(\omega)$
can be associated with longitudinal phonon modes $\bq$ in the heterostructure characterized by coupling matrix elements $g_{\bq}$ and energies $\hbar\Omega_{\bq}$. 
The validity of Eq.~(\ref{eq:propagator}) can be demonstrated by applying it in a well-known limiting case. As we show in the Supporting Information, inserting the 
loss function
into a GW self-energy \cite{hedin_new_1965} yields the carrier-phonon self-energy in random phase approximation \cite{molina-sanchez_temperature-dependent_2016}. Assuming an ideal 2d layer without dielectric embedding then leads to the standard Fröhlich coupling if the layer itself hosts optical phonons.   
For a given $\varepsilon^{\textrm{HS}}_{\bq}(\omega)$ the coupling matrix elements and resonance energies can be extracted from Eq.~(\ref{eq:propagator}), leading to a generalized Fröhlich coupling. A specific example is discussed later on.

The Fröhlich-type matrix elements $g_{\bq}$
define a carrier-phonon interaction Hamiltonian,
\begin{equation}
 \begin{split}
 H_{\textrm{carr-phon}}=\sum_{\bq,\bk,\lambda}g^{\lambda}_{\bq}a^{\dagger}_{\bk,\lambda}a^{\phantom\dagger}_{\bk-\bq,\lambda}\left(b^{\phantom\dagger}_{\bq}+b^{\dagger}_{-\bq}\right)\,,
 \end{split}
\label{eq:H_carr_phon}
\end{equation}
where the phonon modes are characterized by the dispersion $\Omega_{\bq}$. $a^{\dagger}_{\bk,\lambda}$ and $a^{\phantom\dagger}_{\bk,\lambda}$ denote carrier creation and annihilation operators, while $b^{\dagger}_{\bq}$ and $b^{\phantom\dagger}_{\bq}$ represent phonon creation and annihilation operators, respectively. The above Hamiltonian is complemented by a carrier-carrier interaction Hamiltonian that contains Coulomb matrix elements $W^{\textrm{HS,stat}}_{\bq}$ screened by the static part of the heterostructure dielectric function 
with $\varepsilon^{\textrm{s}}(\omega)=\varepsilon_{\infty}$:
\begin{equation}
 \begin{split}
H_{\textrm{Coul}}=\frac{1}{2}\sum_{\bk,\bk',\bq,\lambda\lambda'}W^{\textrm{HS,stat}}_{\bq} a^{\dagger}_{\bk,\lambda}a^{\dagger}_{\bk',\lambda'}a^{\phantom\dagger}_{\bk'-\bq,\lambda'}a^{\phantom\dagger}_{\bk+\bq,\lambda}\,.
\end{split}
\label{eq:H_coul}
\end{equation}
%
Free-carrier and phonon contributions to the Hamiltonian are given by 
\begin{equation}
 \begin{split}
H_0=\sum_{\bk,\lambda}\varepsilon^{\lambda}_{\bk}a^{\dagger}_{\bk,\lambda}a^{\phantom\dagger}_{\bk,\lambda}+\sum_{\bq}\hbar\Omega_{\bq}b^{\dagger}_{\bq}b^{\phantom\dagger}_{\bq}
\end{split}
\label{eq:H_free}
\end{equation}
with band structures $\varepsilon^{\lambda}_{\bk}$. As a result, the total interaction Hamiltonian is reformulated as the sum of a carrier-carrier Coulomb Hamiltonian stemming from the static part of the inverse dielectric function and a carrier-phonon interaction term describing the dynamical part of the inverse dielectric function. 

We evaluate the Heisenberg equation of motion for the above Hamiltonian to derive equations for the microscopic polarizations $\psi_{\bk}(t)=\big\langle a_{\bk,\textrm{c}}^{\dagger}\, a_{\bk,\textrm{v}}^{\phantom\dagger}\big\rangle(t) $, which determine the inter-band optical response of the material. The polarizations contain information about excitonic transitions
if Coulomb interaction is taken into account. \cite{kira_many-body_2006} Introducing two-particle operators $X^{\dagger}_{\nu,\bq}=\sum_{\bk}\phi_{\nu,\bq}(\bk)a_{\bk-\bq,\textrm{c}}^{\dagger}\, a_{\bk,\textrm{v}}^{\phantom\dagger} $ with two-particle wave functions that are solutions of the Wannier equation
\begin{equation}
 \begin{split}
(\varepsilon^{\textrm{c}}_{\bk-\bq}-\varepsilon^{\textrm{v}}_{\bk}-E_{\nu,\bq})\phi_{\nu,\bq}(\bk)-\sum_{\bk'}W^{\textrm{HS,stat}}_{\bk-\bk'}\phi_{\nu,\bq}(\bk')=0\,,
\end{split}
\label{eq:wannier}
\end{equation}
we can eliminate Coulomb interaction from the EOM and directly access excitonic polarizations $\psi_{\nu}(t)=\sum_{\bk}(\phi_{\nu,\boldsymbol{0}}(\bk))^*\psi_{\bk}(t)$. $\bq$ denotes the total momentum of the exciton, while $\nu$ is the exciton quantum number that belongs to the relative motion of electron and hole. If carrier-phonon interaction is described in Born-Markov approximation, we arrive at the EOM \cite{selig_excitonic_2016, krummheuer_theory_2002}:
%
\begin{equation}
 \begin{split}
i\hbar\frac{d}{dt}\psi_{\nu}(t)=E_{\nu,\boldsymbol{0}}\psi_{\nu}(t) +\sum_{\nu'}\big< \nu,\boldsymbol{0} \big| H^{\textrm{eff}} \big| \nu',\boldsymbol{0} \big>\psi_{\nu'}(t)
\end{split}
\label{eq:EOM}
\end{equation}
with the effective Hamiltonian
%
\begin{equation}
 \begin{split}
 &\big< \nu,\boldsymbol{0} \big| H^{\textrm{eff}} \big| \nu',\boldsymbol{0} \big> = \sum_{\alpha\bq}\tilde{G}_{\bq}^{\nu\alpha}(\tilde{G}_{\bq}^{\nu'\alpha})^* \times \\
 &\times\Big(\frac{1+n_{\bq}}{E_{\nu',\boldsymbol{0}}-E_{\alpha,\bq}-\hbar\Omega_{\bq}+i\Gamma } +\frac{n_{\bq}}{E_{\nu',\boldsymbol{0}}-E_{\alpha,\bq}+\hbar\Omega_{\bq}+i\Gamma}\Big)\,.
\end{split}
\label{eq:Heff}
\end{equation}
Here we introduced exciton-phonon matrix elements $\tilde{G}_{\bq}^{\nu\alpha}=\sum_{\bk} (\phi_{\nu,\boldsymbol{0}}(\bk))^*(\phi_{\alpha,\bq}(\bk)g^c_{\bq}-\phi_{\alpha,\bq}(\bk+\bq)g^v_{\bq})$, phonon populations $n_{\bq}$ given by Bose functions and a phenomenological damping $\Gamma$. 
We see that renormalizations of the exciton energies $E_{\nu,\boldsymbol{0}}$ as well as mixture of the bright exciton states $\ket{\nu,\boldsymbol{0}}$ with vanishing total momentum $\bq=\boldsymbol{0} $ are induced by $H^{\textrm{eff}}$. Since $H^{\textrm{eff}}$ itself is not hermitian, we rely on its hermitian part $\mathcal{H} H^{\textrm{eff}}=\frac{1}{2}((H^{\textrm{eff}})^{\dagger}+H^{\textrm{eff}})$.
Then the renormalization of energies can be approximated as a first-order perturbation $\Delta E_{\nu,\boldsymbol{0}}=\big< \nu,\boldsymbol{0} \big|\mathcal{H} H^{\textrm{eff}} \big| \nu,\boldsymbol{0} \big>$, while corrections of exciton wave functions arise from the first-order off-diagonal contribution 
\begin{equation}
 \begin{split}
\Delta \phi_{\nu,\boldsymbol{0}}(\bk)=\sum_{\nu\neq\nu'}\phi_{\nu',\boldsymbol{0}}(\bk)\frac{\big< \nu',\boldsymbol{0} \big|\mathcal{H} H^{\textrm{eff}} \big| \nu,\boldsymbol{0} \big>}{E_{\nu,\boldsymbol{0}}-E_{\nu',\boldsymbol{0}}}\,.
\end{split}
\label{eq:wavef_corr}
\end{equation}
\begin{figure*}[h!t]
\centering
\includegraphics[width=\textwidth]{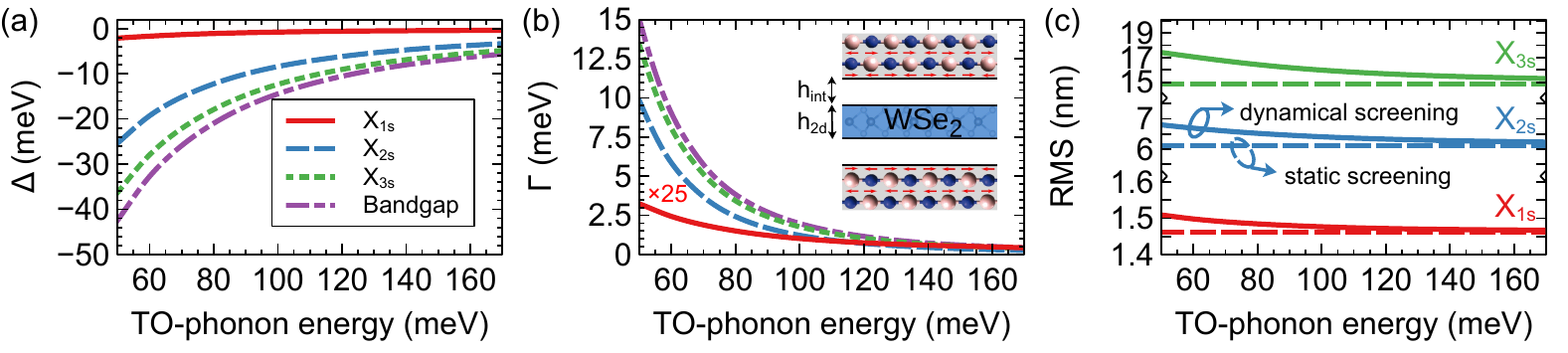}
\caption{Energy renormalizations \textbf{(a)} and linewidth \textbf{(b)} of 1s-, 2s- and 3s-excitons in WSe$_2$ as well as the bandgap transition depending on the energy $\hbar\omega_0$ of the substrate TO-phonon. Both quantities exhibit a characteristic inverse power law dependence.
Note that the binding energy reduction (difference to bandgap) is strongest for $X_{1s}$. The inset shows a schematic of a TMD monolayer coupling to in-plane optical phonons in the surrounding dielectric material, which is detached by an inter-layer gap. The results are obtained for $T=300$ K. The temperature dependence is discussed in the Supporting Information. \textbf{(c)} RMS radius of 1s-, 2s- and 3s-excitons with static and dynamical screening calculated as RMS value of exciton wave functions. As for the energy renormalizations, an inverse power law dependence of the carrier-phonon-induced RMS corrections on $\omega_0$ is obtained.}
\label{fig:renorm}
\label{fig:RMS}
\end{figure*}
For the example of a WSe$_2$ monolayer encapsulated in hBN, we use a macroscopic dielectric function that properly takes into account the heterostructure geometry in terms of nonlocal screening effects. In TMD heterostructures, a finite inter-layer distance (airgap) naturally occurs between the TMD layer 
and the adjacent layers.
~\cite{rooney_observing_2017} In Ref.~\citenum{florian_dielectric_2018}, it has been shown that for this geometry Poisson's equation can be solved analytically including the airgap to obtain the expression
\begin{equation}
 \begin{split}
\varepsilon^{\textrm{HS}}_{\bq}(\omega) = \varepsilon_{\textrm{2d}} \, \frac{1-\epstd_1 \alpha\beta - \epstd_2 \alpha + \epstd_1\epstd_2 \beta}{1+\epstd_1 \alpha\beta + \epstd_2 \alpha + \epstd_1\epstd_2 \beta}~,
 \end{split}
\label{eq:invDF_HS}
\end{equation}
that generalizes the well-known Rytova-Keldysh dielectric function with $\alpha=e^{-q h_{\textrm{2d}}}$, $\beta=e^{-2q h_{\textrm{int}}}$, $\epstd_1=\frac{1-\varepsilon^{\textrm{s}}(\omega)}{1+\varepsilon^{\textrm{s}}(\omega)}$ and $\epstd_2=\frac{\varepsilon_{\textrm{2d}}-1}{\varepsilon_{\textrm{2d}}+1}$. $h_{\textrm{2d}}$ is the thickness of the 2d layer, $h_{\textrm{int}}$ is the inter-layer distance, $\varepsilon_{\textrm{2d}}$ is the dielectric constant of the TMD material and $\varepsilon^{\textrm{s}}(\omega)$ is the substrate dielectric function described by Eq.~\eqref{eq:substrate_DF_TO}. While $\varepsilon_{\textrm{2d}}$ corresponds to the polarizability of the TMD layer itself, $\varepsilon^{\textrm{HS}}_{\bq}(\omega)$ contains the momentum and frequency dependence due to the full heterostructure.

Introducing $~\varepsilon^s(\omega)$ in Eq.~\eqref{eq:invDF_HS}, the inverse of the latter can be 
cast into the form of Eq.~(\ref{eq:propagator}), with exactly two resonances for each substrate phonon mode. Using the 2d Coulomb potential for $V_{\bq}$,
the coupling matrix elements and resonance energies can be determined as: 
\begin{equation}
\begin{split}
 \left|g_{\bq}\right|^2&=\frac{e^2}{2\varepsilon_0 q}\frac{1}{2\Omega_{\bq}}\varepsilon_{\textrm{2d}}\frac{s^2 K}{((c+d\varepsilon_{\infty})\tilde{\varepsilon}^{\textrm{2d}}_{q})^2} \,, \\
 \Omega_{\bq}&=\sqrt{\omega_0^2+\frac{b s^2}{a+b\varepsilon_{\infty}}}\,,
\label{eq:invDF_full}
\end{split}
\end{equation}
with $a=1-\alpha\beta-\epstd_2(\alpha-\beta)$, $b=1+\alpha\beta-\epstd_2(\alpha+\beta)$, $c=1+\alpha\beta+\epstd_2(\alpha+\beta)$, $d=1-\alpha\beta+\epstd_2(\alpha-\beta)$, $K=bc-ad=4(1-\epstd_2^2)\alpha\beta$, $\tilde{\varepsilon}^{\textrm{2d}}_{q}=\varepsilon_{\textrm{2d}}\frac{a+b\varepsilon_{\infty}}{c+d\varepsilon_{\infty}}$. The matrix elements describe a Fröhlich-type coupling of carriers to a long-range polarization generated by optical phonons in the substrate. Note the exponential decay of the coupling strength $\left|g_{\bq}\right|^2\propto K$ with the inter-layer distance $h_{\textrm{int}}$ contained in the parameter $\beta$. This behavior reflects the evanescent nature of the polarization field in vertical direction. The form (\ref{eq:invDF_full}) is maintained for the asymmetric situation of a 2d layer on a substrate without capping, see the Supporting Information.

When applying our approach to hBN, care has to be taken because of the anisotropy of its dielectric response. There are two infrared active phonon modes, the lower-energy one vibrating out of plane (corresponding to $\varepsilon_{\perp}$) and the higher-energy one vibrating in plane (corresponding to $\varepsilon_{\parallel}$). \cite{geick_normal_1966}
As shown by Mele \cite{mele_screening_2001}, in case of an uniaxial anisotropic response with the c-axis as one principal axis, the anisotropy can be described by an effective dielectric constant $\varepsilon_{\textrm{eff}}=\sqrt{\varepsilon_{\parallel}\varepsilon_{\perp}} $. While Mele discusses the case of real-valued dielectric constants, his derivation also holds in case of absorptive, frequency-dependent media. In the effective dielectric function, real and imaginary parts are mixed nontrivially due to the geometric mean. Still, it is possible to approximate the dielectric function by two separate Lorentz oscillators (\ref{eq:substrate_DF_TO}), see the Supporting Information. Using the hBN parameters from Ref.~\citenum{geick_normal_1966}, we obtain $\varepsilon_{\infty}=4.5$, $s_1=68$ meV, $s_2=123$ meV, $\omega_{0,1}=98$ meV and $\omega_{0,2}=172$ meV. 

For the coupling between 2d carriers and environmental phonons
in WSe$_2$, the band structure around the K-point is approximated by electron and hole effective masses $m_e=m_h=0.4m_0$ yielding a reduced exciton mass $\mu_r=0.2m_0$. \cite{stier_magnetooptics_2018} 
We limit carrier-phonon scattering to intra-valley processes neglecting coupling to dark excitons that involve large momentum transfer $q$. \cite{selig_excitonic_2016} This is justified by the $1/q$-scaling behavior of the Fröhlich-type coupling considered here. We use a layer thickness of $h_{\textrm{2d}}=0.66$ nm, a dielectric constant 
$\varepsilon_{\textrm{2d}}=14.6$ corresponding to a screening length $r_0=4.5$ nm \cite{stier_magnetooptics_2018}
and a broadening $\Gamma=10$ meV. For the inter-layer distance, we choose $h_{\textrm{int}}=0.2$ nm.
As fixed parameters of the dielectric environment, the high-frequency dielectric constant of hBN along with the oscillator strength of the low-energy hBN phonon is used. Note that the coupling efficiency (\ref{eq:invDF_full}) approximately scales with $s^2$, so that our results can be easily transferred to different phonon oscillator strengths.

Exciton lineshifts and broadenings due to the interaction with substrate phonons are shown in Fig~\ref{fig:renorm}, as obtained from real and imaginary parts of $\Delta E_{\nu,\boldsymbol{0}}$. The bandgap is defined as the lowest unbound (positive-energy) state from the Wannier equation. The substrate phonon energy $\hbar\omega_0$ is varied between $50$ and $170$ meV to include typical phonon energies of hBN and SiO$_2$. \cite{karimi_dielectric_2016} We find a characteristic inverse power law dependence of renormalization effects on the substrate-phonon energy. Scattering is dominated by the phonon-emission term in the Hamiltonian (\ref{eq:Heff}), which is sensitive to scattering from the state $\ket{\nu,\boldsymbol{0}}$ to states $\ket{\alpha,\bq}$ with energy $E_{\nu,\boldsymbol{0}}-\hbar\Omega_{\bq} $.
In general, phonons of the considered energy range are not matching resonant relaxation processes $\ket{ns}\rightarrow\ket{(n-1)s}$ at low momentum transfer. Hence nonresonant scattering processes dominate, yielding the observed $\omega_0$-dependence as shown in the Supporting Information.
Since for higher exciton states possible final states are energetically more dense, renormalization effects are increasingly efficient.
Prevailing of phonon emission leads to a weak temperature dependence of exciton energy renormalizations, see the Supporting Information.




%
%
From the first-order corrections to exciton wave functions $\Delta \phi_{\nu,\boldsymbol{0}}(\bk)$, we calculate modifications of the exciton root-mean-square (RMS) radius $\textrm{RMS}=\sqrt{\left<\boldsymbol{r}^2\right>-\left<\boldsymbol{r}\right>^2}$ shown in Fig.~\ref{fig:RMS}(c). The systematic increase of the exciton radius due to carrier-phonon interaction is a consequence of binding-energy reduction. 

\begin{figure}[]
\centering
\includegraphics[width=\columnwidth]{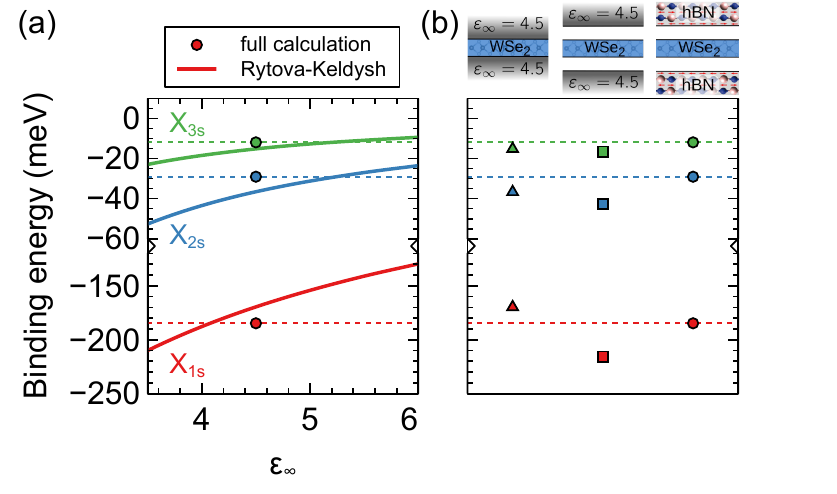}
\caption{\textbf{(a)} Binding energies of 1s-, 2s- and 3s-excitons as a function of static environmental dielectric constant $\varepsilon_{\infty}$ in the absence of an inter-layer gap and without dynamical phonon response. This situation corresponds to a Rytova-Keldysh potential. For comparison, binding energies in the presence of an inter-layer gap of $0.2$ nm, a fixed static constant $\varepsilon_{\infty}=4.5$ for hBN 
as well as additional coupling to hBN phonons 
are given by the filled circles.
The dashed horizontal lines serve as guide to the eye to read off effective dielectric constants for each exciton state.
\textbf{(b)} Exciton binding energies for $\varepsilon_{\infty}=4.5$ obtained at three levels of theory. From left to right: Static screening without inter-layer gap, static screening with inter-layer gap and with hBN-phonon coupling and inter-layer gap.
}
\label{fig:binding}
\end{figure}


Subsequently, we compare our results to frequently used approximations. The full calculation takes into account the interlayer gap and both hBN phonons as described by our double-oscillator model. When a static background screening $\varepsilon_{\infty}$ for the hBN layers is used and the inter-layer air gap between WSe$_2$ and hBN is neglected, the solid lines in Fig.~\ref{fig:binding}(a) are obtained. Clearly, different dielectric constants $\varepsilon_{\infty}$ for various excitonic states are necessary to fit the result of the full calculation as indicated by the horizontal lines. When considering the independently known static high-frequency dielectric constant $\varepsilon_{\infty} = 4.5$ for hBN, a calculation without the interlayer gap leads to smaller (larger) 1s (2s,3s) binding energies, triangles in Fig.~\ref{fig:binding}(b), whereas the correct inclusion of the interlayer air gap results in a significantly larger exciton binding energy in all cases, squares in Fig.~\ref{fig:binding}(b).



Comparing the three cases, we find that dynamical exciton-phonon coupling counteracts the effect of the inter-layer gap.
While each of the effects is significant and physically meaningful, the net result of both effects together is small due to a large degree of compensation. The inter-layer gap leads to weaker static screening and increased exciton binding energies especially for the 1s-exiton. Higher exciton states are less affected due to their larger Bohr radius, as discussed in Ref.~\citenum{florian_dielectric_2018}. On the other hand, the interaction with hBN phonons systematically reduces exciton binding energies since the polaron shift is larger for the band gap than for bound states. The effect is stronger for lower exciton states due to the weaker polaron shifts as shown in Fig.\ref{fig:renorm}(a). This explains why a simple Rytova-Keldysh description of screening by the hBN high-frequency dielectric constant effectively captures the experimentally determined exciton binding energies. \cite{stier_magnetooptics_2018, molas_energy_2019} Nevertheless, the exciton linewidth enhancement as shown in Fig.~\ref{fig:renorm}(b) is only obtained if dynamical coupling to phonons is taken into account and underscores that the dynamical coupling to phonons is a significant effect to the observed exciton spectra. One has to keep in mind that the interaction with intrinsic TMD phonons induces renormalizations on a scale comparable to the effects discussed here. \cite{selig_excitonic_2016} A full theoretical treatment would therefore require to take into account both types of phonons side by side, which is beyond the scope of this paper.




In conclusion, we have derived a Fröhlich-type Hamiltonian describing the coupling of carriers in a 2d material to optical phonons in a surrounding dielectric material. Environmental phonons cause polaron effects well known for carrier-phonon interaction intrinsic to the material itself.
For the specific situation of hBN-encapsulated WSe$_2$ we find that the dynamical dielectric response of the environment reduces exciton binding energies by tens of meV depending on the exciton quantum number.
The presented theory can be used in the active field of 
atomically thin material physics
to augment the available description of exciton interaction with intrinsic phonons. \cite{selig_excitonic_2016, molina-sanchez_ab_2017, selig_dark_2018, merkl_ultrafast_2019}
To experimentally quantify the relative importance of intrinsic and extrinsic phonon effects, we suggest to compare the optical properties of suspended, half-encapsulated and fully encapsulated samples.


\textbf{Acknowledgement}

We acknowledge financial support from the Deutsche Forschungsgemeinschaft (RTG 2247 "Quantum Mechanical Materials Modelling") as well as resources for computational time at the HLRN (Hannover/Berlin).


\begin{thebibliography}{46}%
\makeatletter
\providecommand \@ifxundefined [1]{%
 \@ifx{#1\undefined}
}%
\providecommand \@ifnum [1]{%
 \ifnum #1\expandafter \@firstoftwo
 \else \expandafter \@secondoftwo
 \fi
}%
\providecommand \@ifx [1]{%
 \ifx #1\expandafter \@firstoftwo
 \else \expandafter \@secondoftwo
 \fi
}%
\providecommand \natexlab [1]{#1}%
\providecommand \enquote  [1]{``#1''}%
\providecommand \bibnamefont  [1]{#1}%
\providecommand \bibfnamefont [1]{#1}%
\providecommand \citenamefont [1]{#1}%
\providecommand \href@noop [0]{\@secondoftwo}%
\providecommand \href [0]{\begingroup \@sanitize@url \@href}%
\providecommand \@href[1]{\@@startlink{#1}\@@href}%
\providecommand \@@href[1]{\endgroup#1\@@endlink}%
\providecommand \@sanitize@url [0]{\catcode `\\12\catcode `\$12\catcode
  `\&12\catcode `\#12\catcode `\^12\catcode `\_12\catcode `\%12\relax}%
\providecommand \@@startlink[1]{}%
\providecommand \@@endlink[0]{}%
\providecommand \url  [0]{\begingroup\@sanitize@url \@url }%
\providecommand \@url [1]{\endgroup\@href {#1}{\urlprefix }}%
\providecommand \urlprefix  [0]{URL }%
\providecommand \Eprint [0]{\href }%
\providecommand \doibase [0]{http://dx.doi.org/}%
\providecommand \selectlanguage [0]{\@gobble}%
\providecommand \bibinfo  [0]{\@secondoftwo}%
\providecommand \bibfield  [0]{\@secondoftwo}%
\providecommand \translation [1]{[#1]}%
\providecommand \BibitemOpen [0]{}%
\providecommand \bibitemStop [0]{}%
\providecommand \bibitemNoStop [0]{.\EOS\space}%
\providecommand \EOS [0]{\spacefactor3000\relax}%
\providecommand \BibitemShut  [1]{\csname bibitem#1\endcsname}%
\let\auto@bib@innerbib\@empty
\bibitem [{\citenamefont {Qiu}\ \emph {et~al.}(2013)\citenamefont {Qiu},
  \citenamefont {da~Jornada},\ and\ \citenamefont {Louie}}]{qiu_optical_2013}%
  \BibitemOpen
  \bibfield  {author} {\bibinfo {author} {\bibfnamefont {D.~Y.}\ \bibnamefont
  {Qiu}}, \bibinfo {author} {\bibfnamefont {F.~H.}\ \bibnamefont {da~Jornada}},
  \ and\ \bibinfo {author} {\bibfnamefont {S.~G.}\ \bibnamefont {Louie}},\
  }\href {\doibase 10.1103/PhysRevLett.111.216805} {\bibfield  {journal}
  {\bibinfo  {journal} {Physical Review Letters}\ }\textbf {\bibinfo {volume}
  {111}},\ \bibinfo {pages} {216805} (\bibinfo {year} {2013})}\BibitemShut
  {NoStop}%
\bibitem [{\citenamefont {Chernikov}\ \emph {et~al.}(2014)\citenamefont
  {Chernikov}, \citenamefont {Berkelbach}, \citenamefont {Hill}, \citenamefont
  {Rigosi}, \citenamefont {Li}, \citenamefont {Aslan}, \citenamefont
  {Reichman}, \citenamefont {Hybertsen},\ and\ \citenamefont
  {Heinz}}]{chernikov_exciton_2014}%
  \BibitemOpen
  \bibfield  {author} {\bibinfo {author} {\bibfnamefont {A.}~\bibnamefont
  {Chernikov}}, \bibinfo {author} {\bibfnamefont {T.~C.}\ \bibnamefont
  {Berkelbach}}, \bibinfo {author} {\bibfnamefont {H.~M.}\ \bibnamefont
  {Hill}}, \bibinfo {author} {\bibfnamefont {A.}~\bibnamefont {Rigosi}},
  \bibinfo {author} {\bibfnamefont {Y.}~\bibnamefont {Li}}, \bibinfo {author}
  {\bibfnamefont {O.~B.}\ \bibnamefont {Aslan}}, \bibinfo {author}
  {\bibfnamefont {D.~R.}\ \bibnamefont {Reichman}}, \bibinfo {author}
  {\bibfnamefont {M.~S.}\ \bibnamefont {Hybertsen}}, \ and\ \bibinfo {author}
  {\bibfnamefont {T.~F.}\ \bibnamefont {Heinz}},\ }\href {\doibase
  10.1103/PhysRevLett.113.076802} {\bibfield  {journal} {\bibinfo  {journal}
  {Physical Review Letters}\ }\textbf {\bibinfo {volume} {113}},\ \bibinfo
  {pages} {076802} (\bibinfo {year} {2014})}\BibitemShut {NoStop}%
\bibitem [{\citenamefont {Steinhoff}\ \emph {et~al.}(2014)\citenamefont
  {Steinhoff}, \citenamefont {Rösner}, \citenamefont {Jahnke}, \citenamefont
  {Wehling},\ and\ \citenamefont {Gies}}]{steinhoff_influence_2014}%
  \BibitemOpen
  \bibfield  {author} {\bibinfo {author} {\bibfnamefont {A.}~\bibnamefont
  {Steinhoff}}, \bibinfo {author} {\bibfnamefont {M.}~\bibnamefont {Rösner}},
  \bibinfo {author} {\bibfnamefont {F.}~\bibnamefont {Jahnke}}, \bibinfo
  {author} {\bibfnamefont {T.~O.}\ \bibnamefont {Wehling}}, \ and\ \bibinfo
  {author} {\bibfnamefont {C.}~\bibnamefont {Gies}},\ }\href {\doibase
  10.1021/nl500595u} {\bibfield  {journal} {\bibinfo  {journal} {Nano Letters}\
  }\textbf {\bibinfo {volume} {14}},\ \bibinfo {pages} {3743} (\bibinfo {year}
  {2014})}\BibitemShut {NoStop}%
\bibitem [{\citenamefont {Xu}\ \emph {et~al.}(2014)\citenamefont {Xu},
  \citenamefont {Yao}, \citenamefont {Xiao},\ and\ \citenamefont
  {Heinz}}]{xu_spin_2014}%
  \BibitemOpen
  \bibfield  {author} {\bibinfo {author} {\bibfnamefont {X.}~\bibnamefont
  {Xu}}, \bibinfo {author} {\bibfnamefont {W.}~\bibnamefont {Yao}}, \bibinfo
  {author} {\bibfnamefont {D.}~\bibnamefont {Xiao}}, \ and\ \bibinfo {author}
  {\bibfnamefont {T.~F.}\ \bibnamefont {Heinz}},\ }\href {\doibase
  10.1038/nphys2942} {\bibfield  {journal} {\bibinfo  {journal} {Nature
  Physics}\ }\textbf {\bibinfo {volume} {10}},\ \bibinfo {pages} {343}
  (\bibinfo {year} {2014})}\BibitemShut {NoStop}%
\bibitem [{\citenamefont {Pospischil}\ \emph {et~al.}(2014)\citenamefont
  {Pospischil}, \citenamefont {Furchi},\ and\ \citenamefont
  {Mueller}}]{pospischil_solar-energy_2014}%
  \BibitemOpen
  \bibfield  {author} {\bibinfo {author} {\bibfnamefont {A.}~\bibnamefont
  {Pospischil}}, \bibinfo {author} {\bibfnamefont {M.~M.}\ \bibnamefont
  {Furchi}}, \ and\ \bibinfo {author} {\bibfnamefont {T.}~\bibnamefont
  {Mueller}},\ }\href {\doibase 10.1038/nnano.2014.14} {\bibfield  {journal}
  {\bibinfo  {journal} {Nature Nanotechnology}\ }\textbf {\bibinfo {volume}
  {9}},\ \bibinfo {pages} {257} (\bibinfo {year} {2014})}\BibitemShut {NoStop}%
\bibitem [{\citenamefont {Baugher}\ \emph {et~al.}(2014)\citenamefont
  {Baugher}, \citenamefont {Churchill}, \citenamefont {Yang},\ and\
  \citenamefont {Jarillo-Herrero}}]{baugher_optoelectronic_2014}%
  \BibitemOpen
  \bibfield  {author} {\bibinfo {author} {\bibfnamefont {B.~W.~H.}\
  \bibnamefont {Baugher}}, \bibinfo {author} {\bibfnamefont {H.~O.~H.}\
  \bibnamefont {Churchill}}, \bibinfo {author} {\bibfnamefont {Y.}~\bibnamefont
  {Yang}}, \ and\ \bibinfo {author} {\bibfnamefont {P.}~\bibnamefont
  {Jarillo-Herrero}},\ }\href {\doibase 10.1038/nnano.2014.25} {\bibfield
  {journal} {\bibinfo  {journal} {Nature Nanotechnology}\ }\textbf {\bibinfo
  {volume} {9}},\ \bibinfo {pages} {262} (\bibinfo {year} {2014})}\BibitemShut
  {NoStop}%
\bibitem [{\citenamefont {Ross}\ \emph {et~al.}(2014)\citenamefont {Ross},
  \citenamefont {Klement}, \citenamefont {Jones}, \citenamefont {Ghimire},
  \citenamefont {Yan}, \citenamefont {Mandrus}, \citenamefont {Taniguchi},
  \citenamefont {Watanabe}, \citenamefont {Kitamura}, \citenamefont {Yao},
  \citenamefont {Cobden},\ and\ \citenamefont {Xu}}]{ross_electrically_2014}%
  \BibitemOpen
  \bibfield  {author} {\bibinfo {author} {\bibfnamefont {J.~S.}\ \bibnamefont
  {Ross}}, \bibinfo {author} {\bibfnamefont {P.}~\bibnamefont {Klement}},
  \bibinfo {author} {\bibfnamefont {A.~M.}\ \bibnamefont {Jones}}, \bibinfo
  {author} {\bibfnamefont {N.~J.}\ \bibnamefont {Ghimire}}, \bibinfo {author}
  {\bibfnamefont {J.}~\bibnamefont {Yan}}, \bibinfo {author} {\bibfnamefont
  {D.~G.}\ \bibnamefont {Mandrus}}, \bibinfo {author} {\bibfnamefont
  {T.}~\bibnamefont {Taniguchi}}, \bibinfo {author} {\bibfnamefont
  {K.}~\bibnamefont {Watanabe}}, \bibinfo {author} {\bibfnamefont
  {K.}~\bibnamefont {Kitamura}}, \bibinfo {author} {\bibfnamefont
  {W.}~\bibnamefont {Yao}}, \bibinfo {author} {\bibfnamefont {D.~H.}\
  \bibnamefont {Cobden}}, \ and\ \bibinfo {author} {\bibfnamefont
  {X.}~\bibnamefont {Xu}},\ }\href {http://dx.doi.org/10.1038/nnano.2014.26}
  {\bibfield  {journal} {\bibinfo  {journal} {Nature Nanotechnology}\ }\textbf
  {\bibinfo {volume} {9}},\ \bibinfo {pages} {268} (\bibinfo {year}
  {2014})}\BibitemShut {NoStop}%
\bibitem [{\citenamefont {Withers}\ \emph {et~al.}(2015)\citenamefont
  {Withers}, \citenamefont {Pozo-Zamudio}, \citenamefont {Mishchenko},
  \citenamefont {Rooney}, \citenamefont {Gholinia}, \citenamefont {Watanabe},
  \citenamefont {Taniguchi}, \citenamefont {Haigh}, \citenamefont {Geim},
  \citenamefont {Tartakovskii},\ and\ \citenamefont
  {Novoselov}}]{withers_light-emitting_2015}%
  \BibitemOpen
  \bibfield  {author} {\bibinfo {author} {\bibfnamefont {F.}~\bibnamefont
  {Withers}}, \bibinfo {author} {\bibfnamefont {O.~D.}\ \bibnamefont
  {Pozo-Zamudio}}, \bibinfo {author} {\bibfnamefont {A.}~\bibnamefont
  {Mishchenko}}, \bibinfo {author} {\bibfnamefont {A.~P.}\ \bibnamefont
  {Rooney}}, \bibinfo {author} {\bibfnamefont {A.}~\bibnamefont {Gholinia}},
  \bibinfo {author} {\bibfnamefont {K.}~\bibnamefont {Watanabe}}, \bibinfo
  {author} {\bibfnamefont {T.}~\bibnamefont {Taniguchi}}, \bibinfo {author}
  {\bibfnamefont {S.~J.}\ \bibnamefont {Haigh}}, \bibinfo {author}
  {\bibfnamefont {A.~K.}\ \bibnamefont {Geim}}, \bibinfo {author}
  {\bibfnamefont {A.~I.}\ \bibnamefont {Tartakovskii}}, \ and\ \bibinfo
  {author} {\bibfnamefont {K.~S.}\ \bibnamefont {Novoselov}},\ }\href {\doibase
  10.1038/nmat4205} {\bibfield  {journal} {\bibinfo  {journal} {Nature
  Materials}\ }\textbf {\bibinfo {volume} {14}},\ \bibinfo {pages} {301}
  (\bibinfo {year} {2015})}\BibitemShut {NoStop}%
\bibitem [{\citenamefont {Wu}\ \emph {et~al.}(2015)\citenamefont {Wu},
  \citenamefont {Buckley}, \citenamefont {Schaibley}, \citenamefont {Feng},
  \citenamefont {Yan}, \citenamefont {Mandrus}, \citenamefont {Hatami},
  \citenamefont {Yao}, \citenamefont {Vučković}, \citenamefont {Majumdar},\
  and\ \citenamefont {Xu}}]{wu_monolayer_2015}%
  \BibitemOpen
  \bibfield  {author} {\bibinfo {author} {\bibfnamefont {S.}~\bibnamefont
  {Wu}}, \bibinfo {author} {\bibfnamefont {S.}~\bibnamefont {Buckley}},
  \bibinfo {author} {\bibfnamefont {J.~R.}\ \bibnamefont {Schaibley}}, \bibinfo
  {author} {\bibfnamefont {L.}~\bibnamefont {Feng}}, \bibinfo {author}
  {\bibfnamefont {J.}~\bibnamefont {Yan}}, \bibinfo {author} {\bibfnamefont
  {D.~G.}\ \bibnamefont {Mandrus}}, \bibinfo {author} {\bibfnamefont
  {F.}~\bibnamefont {Hatami}}, \bibinfo {author} {\bibfnamefont
  {W.}~\bibnamefont {Yao}}, \bibinfo {author} {\bibfnamefont {J.}~\bibnamefont
  {Vučković}}, \bibinfo {author} {\bibfnamefont {A.}~\bibnamefont
  {Majumdar}}, \ and\ \bibinfo {author} {\bibfnamefont {X.}~\bibnamefont
  {Xu}},\ }\href {\doibase 10.1038/nature14290} {\bibfield  {journal} {\bibinfo
   {journal} {Nature}\ }\textbf {\bibinfo {volume} {520}},\ \bibinfo {pages}
  {69} (\bibinfo {year} {2015})}\BibitemShut {NoStop}%
\bibitem [{\citenamefont {Ye}\ \emph {et~al.}(2015)\citenamefont {Ye},
  \citenamefont {Wong}, \citenamefont {Lu}, \citenamefont {Ni}, \citenamefont
  {Zhu}, \citenamefont {Chen}, \citenamefont {Wang},\ and\ \citenamefont
  {Zhang}}]{ye_monolayer_2015}%
  \BibitemOpen
  \bibfield  {author} {\bibinfo {author} {\bibfnamefont {Y.}~\bibnamefont
  {Ye}}, \bibinfo {author} {\bibfnamefont {Z.~J.}\ \bibnamefont {Wong}},
  \bibinfo {author} {\bibfnamefont {X.}~\bibnamefont {Lu}}, \bibinfo {author}
  {\bibfnamefont {X.}~\bibnamefont {Ni}}, \bibinfo {author} {\bibfnamefont
  {H.}~\bibnamefont {Zhu}}, \bibinfo {author} {\bibfnamefont {X.}~\bibnamefont
  {Chen}}, \bibinfo {author} {\bibfnamefont {Y.}~\bibnamefont {Wang}}, \ and\
  \bibinfo {author} {\bibfnamefont {X.}~\bibnamefont {Zhang}},\ }\href
  {\doibase 10.1038/nphoton.2015.197} {\bibfield  {journal} {\bibinfo
  {journal} {Nature Photonics}\ }\textbf {\bibinfo {volume} {9}},\ \bibinfo
  {pages} {733–737} (\bibinfo {year} {2015})}\BibitemShut {NoStop}%
\bibitem [{\citenamefont {Salehzadeh}\ \emph {et~al.}(2015)\citenamefont
  {Salehzadeh}, \citenamefont {Djavid}, \citenamefont {Tran}, \citenamefont
  {Shih},\ and\ \citenamefont {Mi}}]{salehzadeh_optically_2015}%
  \BibitemOpen
  \bibfield  {author} {\bibinfo {author} {\bibfnamefont {O.}~\bibnamefont
  {Salehzadeh}}, \bibinfo {author} {\bibfnamefont {M.}~\bibnamefont {Djavid}},
  \bibinfo {author} {\bibfnamefont {N.~H.}\ \bibnamefont {Tran}}, \bibinfo
  {author} {\bibfnamefont {I.}~\bibnamefont {Shih}}, \ and\ \bibinfo {author}
  {\bibfnamefont {Z.}~\bibnamefont {Mi}},\ }\href {\doibase
  10.1021/acs.nanolett.5b01665} {\bibfield  {journal} {\bibinfo  {journal}
  {Nano Letters}\ }\textbf {\bibinfo {volume} {15}},\ \bibinfo {pages} {5302}
  (\bibinfo {year} {2015})},\ \bibinfo {note} {{PMID:} 26214363}\BibitemShut
  {NoStop}%
\bibitem [{\citenamefont {Li}\ \emph {et~al.}(2017)\citenamefont {Li},
  \citenamefont {Zhang}, \citenamefont {Huang}, \citenamefont {Sun},
  \citenamefont {Fan}, \citenamefont {Feng}, \citenamefont {Wang},\ and\
  \citenamefont {Ning}}]{li_room-temperature_2017}%
  \BibitemOpen
  \bibfield  {author} {\bibinfo {author} {\bibfnamefont {Y.}~\bibnamefont
  {Li}}, \bibinfo {author} {\bibfnamefont {J.}~\bibnamefont {Zhang}}, \bibinfo
  {author} {\bibfnamefont {D.}~\bibnamefont {Huang}}, \bibinfo {author}
  {\bibfnamefont {H.}~\bibnamefont {Sun}}, \bibinfo {author} {\bibfnamefont
  {F.}~\bibnamefont {Fan}}, \bibinfo {author} {\bibfnamefont {J.}~\bibnamefont
  {Feng}}, \bibinfo {author} {\bibfnamefont {Z.}~\bibnamefont {Wang}}, \ and\
  \bibinfo {author} {\bibfnamefont {C.~Z.}\ \bibnamefont {Ning}},\ }\href
  {\doibase 10.1038/nnano.2017.128} {\bibfield  {journal} {\bibinfo  {journal}
  {Nature Nanotechnology}\ }\textbf {\bibinfo {volume} {12}},\ \bibinfo {pages}
  {987} (\bibinfo {year} {2017})}\BibitemShut {NoStop}%
\bibitem [{\citenamefont {Geim}\ and\ \citenamefont
  {Grigorieva}(2013)}]{geim_van_2013}%
  \BibitemOpen
  \bibfield  {author} {\bibinfo {author} {\bibfnamefont {A.~K.}\ \bibnamefont
  {Geim}}\ and\ \bibinfo {author} {\bibfnamefont {I.~V.}\ \bibnamefont
  {Grigorieva}},\ }\href {\doibase 10.1038/nature12385} {\bibfield  {journal}
  {\bibinfo  {journal} {Nature}\ }\textbf {\bibinfo {volume} {499}},\ \bibinfo
  {pages} {419} (\bibinfo {year} {2013})}\BibitemShut {NoStop}%
\bibitem [{\citenamefont {Berkelbach}\ \emph {et~al.}(2013)\citenamefont
  {Berkelbach}, \citenamefont {Hybertsen},\ and\ \citenamefont
  {Reichman}}]{berkelbach_theory_2013}%
  \BibitemOpen
  \bibfield  {author} {\bibinfo {author} {\bibfnamefont {T.~C.}\ \bibnamefont
  {Berkelbach}}, \bibinfo {author} {\bibfnamefont {M.~S.}\ \bibnamefont
  {Hybertsen}}, \ and\ \bibinfo {author} {\bibfnamefont {D.~R.}\ \bibnamefont
  {Reichman}},\ }\href {\doibase 10.1103/PhysRevB.88.045318} {\bibfield
  {journal} {\bibinfo  {journal} {Physical Review B}\ }\textbf {\bibinfo
  {volume} {88}},\ \bibinfo {pages} {045318} (\bibinfo {year}
  {2013})}\BibitemShut {NoStop}%
\bibitem [{\citenamefont {Latini}\ \emph {et~al.}(2015)\citenamefont {Latini},
  \citenamefont {Olsen},\ and\ \citenamefont
  {Thygesen}}]{latini_excitons_2015}%
  \BibitemOpen
  \bibfield  {author} {\bibinfo {author} {\bibfnamefont {S.}~\bibnamefont
  {Latini}}, \bibinfo {author} {\bibfnamefont {T.}~\bibnamefont {Olsen}}, \
  and\ \bibinfo {author} {\bibfnamefont {K.~S.}\ \bibnamefont {Thygesen}},\
  }\href {\doibase 10.1103/PhysRevB.92.245123} {\bibfield  {journal} {\bibinfo
  {journal} {Physical Review B}\ }\textbf {\bibinfo {volume} {92}},\ \bibinfo
  {pages} {245123} (\bibinfo {year} {2015})}\BibitemShut {NoStop}%
\bibitem [{\citenamefont {Steinke}\ \emph {et~al.}(2017)\citenamefont
  {Steinke}, \citenamefont {Mourad}, \citenamefont {Rösner}, \citenamefont
  {Lorke}, \citenamefont {Gies}, \citenamefont {Jahnke}, \citenamefont
  {Czycholl},\ and\ \citenamefont {Wehling}}]{steinke_non-invasive_2017}%
  \BibitemOpen
  \bibfield  {author} {\bibinfo {author} {\bibfnamefont {C.}~\bibnamefont
  {Steinke}}, \bibinfo {author} {\bibfnamefont {D.}~\bibnamefont {Mourad}},
  \bibinfo {author} {\bibfnamefont {M.}~\bibnamefont {Rösner}}, \bibinfo
  {author} {\bibfnamefont {M.}~\bibnamefont {Lorke}}, \bibinfo {author}
  {\bibfnamefont {C.}~\bibnamefont {Gies}}, \bibinfo {author} {\bibfnamefont
  {F.}~\bibnamefont {Jahnke}}, \bibinfo {author} {\bibfnamefont
  {G.}~\bibnamefont {Czycholl}}, \ and\ \bibinfo {author} {\bibfnamefont
  {T.~O.}\ \bibnamefont {Wehling}},\ }\href {\doibase
  10.1103/PhysRevB.96.045431} {\bibfield  {journal} {\bibinfo  {journal}
  {Physical Review B}\ }\textbf {\bibinfo {volume} {96}} (\bibinfo {year}
  {2017}),\ 10.1103/PhysRevB.96.045431},\ \bibinfo {note} {{arXiv:}
  1704.06095}\BibitemShut {NoStop}%
\bibitem [{\citenamefont {Trolle}\ \emph {et~al.}(2017)\citenamefont {Trolle},
  \citenamefont {Pedersen},\ and\ \citenamefont
  {Véniard}}]{trolle_model_2017}%
  \BibitemOpen
  \bibfield  {author} {\bibinfo {author} {\bibfnamefont {M.~L.}\ \bibnamefont
  {Trolle}}, \bibinfo {author} {\bibfnamefont {T.~G.}\ \bibnamefont
  {Pedersen}}, \ and\ \bibinfo {author} {\bibfnamefont {V.}~\bibnamefont
  {Véniard}},\ }\href {\doibase 10.1038/srep39844} {\bibfield  {journal}
  {\bibinfo  {journal} {Scientific Reports}\ }\textbf {\bibinfo {volume} {7}},\
  \bibinfo {pages} {39844} (\bibinfo {year} {2017})}\BibitemShut {NoStop}%
\bibitem [{\citenamefont {Raja}\ \emph {et~al.}(2017)\citenamefont {Raja},
  \citenamefont {Chaves}, \citenamefont {Yu}, \citenamefont {Arefe},
  \citenamefont {Hill}, \citenamefont {Rigosi}, \citenamefont {Berkelbach},
  \citenamefont {Nagler}, \citenamefont {Schüller}, \citenamefont {Korn},
  \citenamefont {Nuckolls}, \citenamefont {Hone}, \citenamefont {Brus},
  \citenamefont {Heinz}, \citenamefont {Reichman},\ and\ \citenamefont
  {Chernikov}}]{raja_coulomb_2017}%
  \BibitemOpen
  \bibfield  {author} {\bibinfo {author} {\bibfnamefont {A.}~\bibnamefont
  {Raja}}, \bibinfo {author} {\bibfnamefont {A.}~\bibnamefont {Chaves}},
  \bibinfo {author} {\bibfnamefont {J.}~\bibnamefont {Yu}}, \bibinfo {author}
  {\bibfnamefont {G.}~\bibnamefont {Arefe}}, \bibinfo {author} {\bibfnamefont
  {H.~M.}\ \bibnamefont {Hill}}, \bibinfo {author} {\bibfnamefont {A.~F.}\
  \bibnamefont {Rigosi}}, \bibinfo {author} {\bibfnamefont {T.~C.}\
  \bibnamefont {Berkelbach}}, \bibinfo {author} {\bibfnamefont
  {P.}~\bibnamefont {Nagler}}, \bibinfo {author} {\bibfnamefont
  {C.}~\bibnamefont {Schüller}}, \bibinfo {author} {\bibfnamefont
  {T.}~\bibnamefont {Korn}}, \bibinfo {author} {\bibfnamefont {C.}~\bibnamefont
  {Nuckolls}}, \bibinfo {author} {\bibfnamefont {J.}~\bibnamefont {Hone}},
  \bibinfo {author} {\bibfnamefont {L.~E.}\ \bibnamefont {Brus}}, \bibinfo
  {author} {\bibfnamefont {T.~F.}\ \bibnamefont {Heinz}}, \bibinfo {author}
  {\bibfnamefont {D.~R.}\ \bibnamefont {Reichman}}, \ and\ \bibinfo {author}
  {\bibfnamefont {A.}~\bibnamefont {Chernikov}},\ }\href {\doibase
  10.1038/ncomms15251} {\bibfield  {journal} {\bibinfo  {journal} {Nature
  Communications}\ }\textbf {\bibinfo {volume} {8}},\ \bibinfo {pages} {15251}
  (\bibinfo {year} {2017})}\BibitemShut {NoStop}%
\bibitem [{\citenamefont {Steinhoff}\ \emph {et~al.}(2017)\citenamefont
  {Steinhoff}, \citenamefont {Florian}, \citenamefont {Rösner}, \citenamefont
  {Schönhoff}, \citenamefont {Wehling},\ and\ \citenamefont
  {Jahnke}}]{steinhoff_exciton_2017}%
  \BibitemOpen
  \bibfield  {author} {\bibinfo {author} {\bibfnamefont {A.}~\bibnamefont
  {Steinhoff}}, \bibinfo {author} {\bibfnamefont {M.}~\bibnamefont {Florian}},
  \bibinfo {author} {\bibfnamefont {M.}~\bibnamefont {Rösner}}, \bibinfo
  {author} {\bibfnamefont {G.}~\bibnamefont {Schönhoff}}, \bibinfo {author}
  {\bibfnamefont {T.~O.}\ \bibnamefont {Wehling}}, \ and\ \bibinfo {author}
  {\bibfnamefont {F.}~\bibnamefont {Jahnke}},\ }\href {\doibase
  10.1038/s41467-017-01298-6} {\bibfield  {journal} {\bibinfo  {journal}
  {Nature Communications}\ }\textbf {\bibinfo {volume} {8}},\ \bibinfo {pages}
  {1166} (\bibinfo {year} {2017})}\BibitemShut {NoStop}%
\bibitem [{\citenamefont {Meckbach}\ \emph
  {et~al.}(2018{\natexlab{a}})\citenamefont {Meckbach}, \citenamefont
  {Stroucken},\ and\ \citenamefont {Koch}}]{meckbach_influence_2018}%
  \BibitemOpen
  \bibfield  {author} {\bibinfo {author} {\bibfnamefont {L.}~\bibnamefont
  {Meckbach}}, \bibinfo {author} {\bibfnamefont {T.}~\bibnamefont {Stroucken}},
  \ and\ \bibinfo {author} {\bibfnamefont {S.~W.}\ \bibnamefont {Koch}},\
  }\href {\doibase 10.1103/PhysRevB.97.035425} {\bibfield  {journal} {\bibinfo
  {journal} {Physical Review B}\ }\textbf {\bibinfo {volume} {97}},\ \bibinfo
  {pages} {035425} (\bibinfo {year} {2018}{\natexlab{a}})}\BibitemShut
  {NoStop}%
\bibitem [{\citenamefont {Florian}\ \emph {et~al.}(2018)\citenamefont
  {Florian}, \citenamefont {Hartmann}, \citenamefont {Steinhoff}, \citenamefont
  {Klein}, \citenamefont {Holleitner}, \citenamefont {Finley}, \citenamefont
  {Wehling}, \citenamefont {Kaniber},\ and\ \citenamefont
  {Gies}}]{florian_dielectric_2018}%
  \BibitemOpen
  \bibfield  {author} {\bibinfo {author} {\bibfnamefont {M.}~\bibnamefont
  {Florian}}, \bibinfo {author} {\bibfnamefont {M.}~\bibnamefont {Hartmann}},
  \bibinfo {author} {\bibfnamefont {A.}~\bibnamefont {Steinhoff}}, \bibinfo
  {author} {\bibfnamefont {J.}~\bibnamefont {Klein}}, \bibinfo {author}
  {\bibfnamefont {A.~W.}\ \bibnamefont {Holleitner}}, \bibinfo {author}
  {\bibfnamefont {J.~J.}\ \bibnamefont {Finley}}, \bibinfo {author}
  {\bibfnamefont {T.~O.}\ \bibnamefont {Wehling}}, \bibinfo {author}
  {\bibfnamefont {M.}~\bibnamefont {Kaniber}}, \ and\ \bibinfo {author}
  {\bibfnamefont {C.}~\bibnamefont {Gies}},\ }\href {\doibase
  10.1021/acs.nanolett.8b00840} {\bibfield  {journal} {\bibinfo  {journal}
  {Nano Letters}\ }\textbf {\bibinfo {volume} {18}},\ \bibinfo {pages} {2725}
  (\bibinfo {year} {2018})}\BibitemShut {NoStop}%
\bibitem [{\citenamefont {Meckbach}\ \emph
  {et~al.}(2018{\natexlab{b}})\citenamefont {Meckbach}, \citenamefont
  {Stroucken},\ and\ \citenamefont {Koch}}]{meckbach_giant_2018}%
  \BibitemOpen
  \bibfield  {author} {\bibinfo {author} {\bibfnamefont {L.}~\bibnamefont
  {Meckbach}}, \bibinfo {author} {\bibfnamefont {T.}~\bibnamefont {Stroucken}},
  \ and\ \bibinfo {author} {\bibfnamefont {S.~W.}\ \bibnamefont {Koch}},\
  }\href {\doibase 10.1063/1.5017069} {\bibfield  {journal} {\bibinfo
  {journal} {Applied Physics Letters}\ }\textbf {\bibinfo {volume} {112}},\
  \bibinfo {pages} {061104} (\bibinfo {year} {2018}{\natexlab{b}})}\BibitemShut
  {NoStop}%
\bibitem [{\citenamefont {Cadiz}\ \emph {et~al.}(2017)\citenamefont {Cadiz},
  \citenamefont {Courtade}, \citenamefont {Robert}, \citenamefont {Wang},
  \citenamefont {Shen}, \citenamefont {Cai}, \citenamefont {Taniguchi},
  \citenamefont {Watanabe}, \citenamefont {Carrere}, \citenamefont {Lagarde},
  \citenamefont {Manca}, \citenamefont {Amand}, \citenamefont {Renucci},
  \citenamefont {Tongay}, \citenamefont {Marie},\ and\ \citenamefont
  {Urbaszek}}]{cadiz_excitonic_2017}%
  \BibitemOpen
  \bibfield  {author} {\bibinfo {author} {\bibfnamefont {F.}~\bibnamefont
  {Cadiz}}, \bibinfo {author} {\bibfnamefont {E.}~\bibnamefont {Courtade}},
  \bibinfo {author} {\bibfnamefont {C.}~\bibnamefont {Robert}}, \bibinfo
  {author} {\bibfnamefont {G.}~\bibnamefont {Wang}}, \bibinfo {author}
  {\bibfnamefont {Y.}~\bibnamefont {Shen}}, \bibinfo {author} {\bibfnamefont
  {H.}~\bibnamefont {Cai}}, \bibinfo {author} {\bibfnamefont {T.}~\bibnamefont
  {Taniguchi}}, \bibinfo {author} {\bibfnamefont {K.}~\bibnamefont {Watanabe}},
  \bibinfo {author} {\bibfnamefont {H.}~\bibnamefont {Carrere}}, \bibinfo
  {author} {\bibfnamefont {D.}~\bibnamefont {Lagarde}}, \bibinfo {author}
  {\bibfnamefont {M.}~\bibnamefont {Manca}}, \bibinfo {author} {\bibfnamefont
  {T.}~\bibnamefont {Amand}}, \bibinfo {author} {\bibfnamefont
  {P.}~\bibnamefont {Renucci}}, \bibinfo {author} {\bibfnamefont
  {S.}~\bibnamefont {Tongay}}, \bibinfo {author} {\bibfnamefont
  {X.}~\bibnamefont {Marie}}, \ and\ \bibinfo {author} {\bibfnamefont
  {B.}~\bibnamefont {Urbaszek}},\ }\href {\doibase 10.1103/PhysRevX.7.021026}
  {\bibfield  {journal} {\bibinfo  {journal} {Physical Review X}\ }\textbf
  {\bibinfo {volume} {7}},\ \bibinfo {pages} {021026} (\bibinfo {year}
  {2017})}\BibitemShut {NoStop}%
\bibitem [{\citenamefont {Keldysh}(1979)}]{keldysh_coulomb_1979}%
  \BibitemOpen
  \bibfield  {author} {\bibinfo {author} {\bibfnamefont {L.~V.}\ \bibnamefont
  {Keldysh}},\ }\href {http://adsabs.harvard.edu/abs/1979JETPL..29..658K}
  {\bibfield  {journal} {\bibinfo  {journal} {Soviet Journal of Experimental
  and Theoretical Physics Letters}\ }\textbf {\bibinfo {volume} {29}},\
  \bibinfo {pages} {658} (\bibinfo {year} {1979})}\BibitemShut {NoStop}%
\bibitem [{\citenamefont {Rytova}(2018)}]{rytova_screened_2018}%
  \BibitemOpen
  \bibfield  {author} {\bibinfo {author} {\bibfnamefont {N.~S.}\ \bibnamefont
  {Rytova}},\ }\href {http://arxiv.org/abs/1806.00976} {\bibfield  {journal}
  {\bibinfo  {journal} {{arXiv:1806.00976} [cond-mat]}\ } (\bibinfo {year}
  {2018})},\ \bibinfo {note} {{arXiv:} 1806.00976}\BibitemShut {NoStop}%
\bibitem [{\citenamefont {Bechstedt}\ and\ \citenamefont
  {Furthmüller}(2019)}]{bechstedt_influence_2019}%
  \BibitemOpen
  \bibfield  {author} {\bibinfo {author} {\bibfnamefont {F.}~\bibnamefont
  {Bechstedt}}\ and\ \bibinfo {author} {\bibfnamefont {J.}~\bibnamefont
  {Furthmüller}},\ }\href {\doibase 10.1063/1.5084324} {\bibfield  {journal}
  {\bibinfo  {journal} {Applied Physics Letters}\ }\textbf {\bibinfo {volume}
  {114}},\ \bibinfo {pages} {122101} (\bibinfo {year} {2019})}\BibitemShut
  {NoStop}%
\bibitem [{\citenamefont {Karimi}\ \emph {et~al.}(2016)\citenamefont {Karimi},
  \citenamefont {Davoody},\ and\ \citenamefont
  {Knezevic}}]{karimi_dielectric_2016}%
  \BibitemOpen
  \bibfield  {author} {\bibinfo {author} {\bibfnamefont {F.}~\bibnamefont
  {Karimi}}, \bibinfo {author} {\bibfnamefont {A.~H.}\ \bibnamefont {Davoody}},
  \ and\ \bibinfo {author} {\bibfnamefont {I.}~\bibnamefont {Knezevic}},\
  }\href {\doibase 10.1103/PhysRevB.93.205421} {\bibfield  {journal} {\bibinfo
  {journal} {Physical Review B}\ }\textbf {\bibinfo {volume} {93}},\ \bibinfo
  {pages} {205421} (\bibinfo {year} {2016})}\BibitemShut {NoStop}%
\bibitem [{\citenamefont {Hwang}\ \emph {et~al.}(2010)\citenamefont {Hwang},
  \citenamefont {Sensarma},\ and\ \citenamefont
  {Das~Sarma}}]{hwang_plasmon-phonon_2010}%
  \BibitemOpen
  \bibfield  {author} {\bibinfo {author} {\bibfnamefont {E.~H.}\ \bibnamefont
  {Hwang}}, \bibinfo {author} {\bibfnamefont {R.}~\bibnamefont {Sensarma}}, \
  and\ \bibinfo {author} {\bibfnamefont {S.}~\bibnamefont {Das~Sarma}},\ }\href
  {\doibase 10.1103/PhysRevB.82.195406} {\bibfield  {journal} {\bibinfo
  {journal} {Physical Review B}\ }\textbf {\bibinfo {volume} {82}},\ \bibinfo
  {pages} {195406} (\bibinfo {year} {2010})}\BibitemShut {NoStop}%
\bibitem [{\citenamefont {Steinhoff}\ \emph {et~al.}(2018)\citenamefont
  {Steinhoff}, \citenamefont {Wehling},\ and\ \citenamefont
  {Rösner}}]{steinhoff_frequency-dependent_2018}%
  \BibitemOpen
  \bibfield  {author} {\bibinfo {author} {\bibfnamefont {A.}~\bibnamefont
  {Steinhoff}}, \bibinfo {author} {\bibfnamefont {T.~O.}\ \bibnamefont
  {Wehling}}, \ and\ \bibinfo {author} {\bibfnamefont {M.}~\bibnamefont
  {Rösner}},\ }\href {\doibase 10.1103/PhysRevB.98.045304} {\bibfield
  {journal} {\bibinfo  {journal} {Physical Review B}\ }\textbf {\bibinfo
  {volume} {98}},\ \bibinfo {pages} {045304} (\bibinfo {year}
  {2018})}\BibitemShut {NoStop}%
\bibitem [{\citenamefont {Chow}\ \emph {et~al.}(2017)\citenamefont {Chow},
  \citenamefont {Yu}, \citenamefont {Jones}, \citenamefont {Yan}, \citenamefont
  {Mandrus}, \citenamefont {Taniguchi}, \citenamefont {Watanabe}, \citenamefont
  {Yao},\ and\ \citenamefont {Xu}}]{chow_unusual_2017}%
  \BibitemOpen
  \bibfield  {author} {\bibinfo {author} {\bibfnamefont {C.~M.}\ \bibnamefont
  {Chow}}, \bibinfo {author} {\bibfnamefont {H.}~\bibnamefont {Yu}}, \bibinfo
  {author} {\bibfnamefont {A.~M.}\ \bibnamefont {Jones}}, \bibinfo {author}
  {\bibfnamefont {J.}~\bibnamefont {Yan}}, \bibinfo {author} {\bibfnamefont
  {D.~G.}\ \bibnamefont {Mandrus}}, \bibinfo {author} {\bibfnamefont
  {T.}~\bibnamefont {Taniguchi}}, \bibinfo {author} {\bibfnamefont
  {K.}~\bibnamefont {Watanabe}}, \bibinfo {author} {\bibfnamefont
  {W.}~\bibnamefont {Yao}}, \ and\ \bibinfo {author} {\bibfnamefont
  {X.}~\bibnamefont {Xu}},\ }\href {\doibase 10.1021/acs.nanolett.6b04944}
  {\bibfield  {journal} {\bibinfo  {journal} {Nano Letters}\ }\textbf {\bibinfo
  {volume} {17}},\ \bibinfo {pages} {1194} (\bibinfo {year}
  {2017})}\BibitemShut {NoStop}%
\bibitem [{\citenamefont {Jin}\ \emph {et~al.}(2017)\citenamefont {Jin},
  \citenamefont {Kim}, \citenamefont {Suh}, \citenamefont {Shi}, \citenamefont
  {Chen}, \citenamefont {Fan}, \citenamefont {Kam}, \citenamefont {Watanabe},
  \citenamefont {Taniguchi}, \citenamefont {Tongay}, \citenamefont {Zettl},
  \citenamefont {Wu},\ and\ \citenamefont {Wang}}]{jin_interlayer_2017}%
  \BibitemOpen
  \bibfield  {author} {\bibinfo {author} {\bibfnamefont {C.}~\bibnamefont
  {Jin}}, \bibinfo {author} {\bibfnamefont {J.}~\bibnamefont {Kim}}, \bibinfo
  {author} {\bibfnamefont {J.}~\bibnamefont {Suh}}, \bibinfo {author}
  {\bibfnamefont {Z.}~\bibnamefont {Shi}}, \bibinfo {author} {\bibfnamefont
  {B.}~\bibnamefont {Chen}}, \bibinfo {author} {\bibfnamefont {X.}~\bibnamefont
  {Fan}}, \bibinfo {author} {\bibfnamefont {M.}~\bibnamefont {Kam}}, \bibinfo
  {author} {\bibfnamefont {K.}~\bibnamefont {Watanabe}}, \bibinfo {author}
  {\bibfnamefont {T.}~\bibnamefont {Taniguchi}}, \bibinfo {author}
  {\bibfnamefont {S.}~\bibnamefont {Tongay}}, \bibinfo {author} {\bibfnamefont
  {A.}~\bibnamefont {Zettl}}, \bibinfo {author} {\bibfnamefont
  {J.}~\bibnamefont {Wu}}, \ and\ \bibinfo {author} {\bibfnamefont
  {F.}~\bibnamefont {Wang}},\ }\href {\doibase 10.1038/nphys3928} {\bibfield
  {journal} {\bibinfo  {journal} {Nature Physics}\ }\textbf {\bibinfo {volume}
  {13}},\ \bibinfo {pages} {127} (\bibinfo {year} {2017})}\BibitemShut
  {NoStop}%
\bibitem [{\citenamefont {Selig}\ \emph {et~al.}(2016)\citenamefont {Selig},
  \citenamefont {Berghäuser}, \citenamefont {Raja}, \citenamefont {Nagler},
  \citenamefont {Schüller}, \citenamefont {Heinz}, \citenamefont {Korn},
  \citenamefont {Chernikov}, \citenamefont {Malic},\ and\ \citenamefont
  {Knorr}}]{selig_excitonic_2016}%
  \BibitemOpen
  \bibfield  {author} {\bibinfo {author} {\bibfnamefont {M.}~\bibnamefont
  {Selig}}, \bibinfo {author} {\bibfnamefont {G.}~\bibnamefont {Berghäuser}},
  \bibinfo {author} {\bibfnamefont {A.}~\bibnamefont {Raja}}, \bibinfo {author}
  {\bibfnamefont {P.}~\bibnamefont {Nagler}}, \bibinfo {author} {\bibfnamefont
  {C.}~\bibnamefont {Schüller}}, \bibinfo {author} {\bibfnamefont {T.~F.}\
  \bibnamefont {Heinz}}, \bibinfo {author} {\bibfnamefont {T.}~\bibnamefont
  {Korn}}, \bibinfo {author} {\bibfnamefont {A.}~\bibnamefont {Chernikov}},
  \bibinfo {author} {\bibfnamefont {E.}~\bibnamefont {Malic}}, \ and\ \bibinfo
  {author} {\bibfnamefont {A.}~\bibnamefont {Knorr}},\ }\href {\doibase
  10.1038/ncomms13279} {\bibfield  {journal} {\bibinfo  {journal} {Nature
  Communications}\ }\textbf {\bibinfo {volume} {7}},\ \bibinfo {pages} {13279}
  (\bibinfo {year} {2016})}\BibitemShut {NoStop}%
\bibitem [{\citenamefont {Geick}\ \emph {et~al.}(1966)\citenamefont {Geick},
  \citenamefont {Perry},\ and\ \citenamefont {Rupprecht}}]{geick_normal_1966}%
  \BibitemOpen
  \bibfield  {author} {\bibinfo {author} {\bibfnamefont {R.}~\bibnamefont
  {Geick}}, \bibinfo {author} {\bibfnamefont {C.~H.}\ \bibnamefont {Perry}}, \
  and\ \bibinfo {author} {\bibfnamefont {G.}~\bibnamefont {Rupprecht}},\ }\href
  {\doibase 10.1103/PhysRev.146.543} {\bibfield  {journal} {\bibinfo  {journal}
  {Physical Review}\ }\textbf {\bibinfo {volume} {146}},\ \bibinfo {pages}
  {543} (\bibinfo {year} {1966})}\BibitemShut {NoStop}%
\bibitem [{\citenamefont {Haug}\ and\ \citenamefont
  {Schmitt-Rink}(1984)}]{haug_electron_1984}%
  \BibitemOpen
  \bibfield  {author} {\bibinfo {author} {\bibfnamefont {H.}~\bibnamefont
  {Haug}}\ and\ \bibinfo {author} {\bibfnamefont {S.}~\bibnamefont
  {Schmitt-Rink}},\ }\href {\doibase 10.1016/0079-6727(84)90026-0} {\bibfield
  {journal} {\bibinfo  {journal} {Progress in Quantum Electronics}\ }\textbf
  {\bibinfo {volume} {9}},\ \bibinfo {pages} {3} (\bibinfo {year}
  {1984})}\BibitemShut {NoStop}%
\bibitem [{\citenamefont {Leeuwen}(2004)}]{leeuwen_first-principles_2004}%
  \BibitemOpen
  \bibfield  {author} {\bibinfo {author} {\bibfnamefont {R.~v.}\ \bibnamefont
  {Leeuwen}},\ }\href {\doibase 10.1103/PhysRevB.69.199901} {\bibfield
  {journal} {\bibinfo  {journal} {Phys. Rev. B}\ }\textbf {\bibinfo {volume}
  {69}},\ \bibinfo {pages} {115110} (\bibinfo {year} {2004})}\BibitemShut
  {NoStop}%
\bibitem [{\citenamefont {Hedin}(1965)}]{hedin_new_1965}%
  \BibitemOpen
  \bibfield  {author} {\bibinfo {author} {\bibfnamefont {L.}~\bibnamefont
  {Hedin}},\ }\href {\doibase 10.1103/PhysRev.139.A796} {\bibfield  {journal}
  {\bibinfo  {journal} {Physical Review}\ }\textbf {\bibinfo {volume} {139}},\
  \bibinfo {pages} {A796} (\bibinfo {year} {1965})}\BibitemShut {NoStop}%
\bibitem [{\citenamefont {Molina-Sánchez}\ \emph {et~al.}(2016)\citenamefont
  {Molina-Sánchez}, \citenamefont {Palummo}, \citenamefont {Marini},\ and\
  \citenamefont {Wirtz}}]{molina-sanchez_temperature-dependent_2016}%
  \BibitemOpen
  \bibfield  {author} {\bibinfo {author} {\bibfnamefont {A.}~\bibnamefont
  {Molina-Sánchez}}, \bibinfo {author} {\bibfnamefont {M.}~\bibnamefont
  {Palummo}}, \bibinfo {author} {\bibfnamefont {A.}~\bibnamefont {Marini}}, \
  and\ \bibinfo {author} {\bibfnamefont {L.}~\bibnamefont {Wirtz}},\ }\href
  {\doibase 10.1103/PhysRevB.93.155435} {\bibfield  {journal} {\bibinfo
  {journal} {Physical Review B}\ }\textbf {\bibinfo {volume} {93}},\ \bibinfo
  {pages} {155435} (\bibinfo {year} {2016})}\BibitemShut {NoStop}%
\bibitem [{\citenamefont {Kira}\ and\ \citenamefont
  {Koch}(2006)}]{kira_many-body_2006}%
  \BibitemOpen
  \bibfield  {author} {\bibinfo {author} {\bibfnamefont {M.}~\bibnamefont
  {Kira}}\ and\ \bibinfo {author} {\bibfnamefont {S.~W.}\ \bibnamefont
  {Koch}},\ }\href {\doibase 10.1016/j.pquantelec.2006.12.002} {\bibfield
  {journal} {\bibinfo  {journal} {Progress in Quantum Electronics}\ }\textbf
  {\bibinfo {volume} {30}},\ \bibinfo {pages} {155} (\bibinfo {year}
  {2006})}\BibitemShut {NoStop}%
\bibitem [{\citenamefont {Krummheuer}\ \emph {et~al.}(2002)\citenamefont
  {Krummheuer}, \citenamefont {Axt},\ and\ \citenamefont
  {Kuhn}}]{krummheuer_theory_2002}%
  \BibitemOpen
  \bibfield  {author} {\bibinfo {author} {\bibfnamefont {B.}~\bibnamefont
  {Krummheuer}}, \bibinfo {author} {\bibfnamefont {V.~M.}\ \bibnamefont {Axt}},
  \ and\ \bibinfo {author} {\bibfnamefont {T.}~\bibnamefont {Kuhn}},\ }\href
  {\doibase 10.1103/PhysRevB.65.195313} {\bibfield  {journal} {\bibinfo
  {journal} {Physical Review B}\ }\textbf {\bibinfo {volume} {65}},\ \bibinfo
  {pages} {195313} (\bibinfo {year} {2002})}\BibitemShut {NoStop}%
\bibitem [{\citenamefont {Rooney}\ \emph {et~al.}(2017)\citenamefont {Rooney},
  \citenamefont {Kozikov}, \citenamefont {Rudenko}, \citenamefont {Prestat},
  \citenamefont {Hamer}, \citenamefont {Withers}, \citenamefont {Cao},
  \citenamefont {Novoselov}, \citenamefont {Katsnelson}, \citenamefont
  {Gorbachev},\ and\ \citenamefont {Haigh}}]{rooney_observing_2017}%
  \BibitemOpen
  \bibfield  {author} {\bibinfo {author} {\bibfnamefont {A.~P.}\ \bibnamefont
  {Rooney}}, \bibinfo {author} {\bibfnamefont {A.}~\bibnamefont {Kozikov}},
  \bibinfo {author} {\bibfnamefont {A.~N.}\ \bibnamefont {Rudenko}}, \bibinfo
  {author} {\bibfnamefont {E.}~\bibnamefont {Prestat}}, \bibinfo {author}
  {\bibfnamefont {M.~J.}\ \bibnamefont {Hamer}}, \bibinfo {author}
  {\bibfnamefont {F.}~\bibnamefont {Withers}}, \bibinfo {author} {\bibfnamefont
  {Y.}~\bibnamefont {Cao}}, \bibinfo {author} {\bibfnamefont {K.~S.}\
  \bibnamefont {Novoselov}}, \bibinfo {author} {\bibfnamefont {M.~I.}\
  \bibnamefont {Katsnelson}}, \bibinfo {author} {\bibfnamefont
  {R.}~\bibnamefont {Gorbachev}}, \ and\ \bibinfo {author} {\bibfnamefont
  {S.~J.}\ \bibnamefont {Haigh}},\ }\href {\doibase
  10.1021/acs.nanolett.7b01248} {\bibfield  {journal} {\bibinfo  {journal}
  {Nano Letters}\ } (\bibinfo {year} {2017}),\
  10.1021/acs.nanolett.7b01248}\BibitemShut {NoStop}%
\bibitem [{\citenamefont {Mele}(2001)}]{mele_screening_2001}%
  \BibitemOpen
  \bibfield  {author} {\bibinfo {author} {\bibfnamefont {E.~J.}\ \bibnamefont
  {Mele}},\ }\href {\doibase 10.1119/1.1341252} {\bibfield  {journal} {\bibinfo
   {journal} {American Journal of Physics}\ }\textbf {\bibinfo {volume} {69}},\
  \bibinfo {pages} {557} (\bibinfo {year} {2001})}\BibitemShut {NoStop}%
\bibitem [{\citenamefont {Stier}\ \emph {et~al.}(2018)\citenamefont {Stier},
  \citenamefont {Wilson}, \citenamefont {Velizhanin}, \citenamefont {Kono},
  \citenamefont {Xu},\ and\ \citenamefont
  {Crooker}}]{stier_magnetooptics_2018}%
  \BibitemOpen
  \bibfield  {author} {\bibinfo {author} {\bibfnamefont {A.}~\bibnamefont
  {Stier}}, \bibinfo {author} {\bibfnamefont {N.}~\bibnamefont {Wilson}},
  \bibinfo {author} {\bibfnamefont {K.}~\bibnamefont {Velizhanin}}, \bibinfo
  {author} {\bibfnamefont {J.}~\bibnamefont {Kono}}, \bibinfo {author}
  {\bibfnamefont {X.}~\bibnamefont {Xu}}, \ and\ \bibinfo {author}
  {\bibfnamefont {S.}~\bibnamefont {Crooker}},\ }\href {\doibase
  10.1103/PhysRevLett.120.057405} {\bibfield  {journal} {\bibinfo  {journal}
  {Physical Review Letters}\ }\textbf {\bibinfo {volume} {120}},\ \bibinfo
  {pages} {057405} (\bibinfo {year} {2018})}\BibitemShut {NoStop}%
\bibitem [{\citenamefont {Molas}\ \emph {et~al.}(2019)\citenamefont {Molas},
  \citenamefont {Slobodeniuk}, \citenamefont {Nogajewski}, \citenamefont
  {Bartos}, \citenamefont {Bala}, \citenamefont {Babiński}, \citenamefont
  {Watanabe}, \citenamefont {Taniguchi}, \citenamefont {Faugeras},\ and\
  \citenamefont {Potemski}}]{molas_energy_2019}%
  \BibitemOpen
  \bibfield  {author} {\bibinfo {author} {\bibfnamefont {M.~R.}\ \bibnamefont
  {Molas}}, \bibinfo {author} {\bibfnamefont {A.~O.}\ \bibnamefont
  {Slobodeniuk}}, \bibinfo {author} {\bibfnamefont {K.}~\bibnamefont
  {Nogajewski}}, \bibinfo {author} {\bibfnamefont {M.}~\bibnamefont {Bartos}},
  \bibinfo {author} {\bibfnamefont {{\L}.}~\bibnamefont {Bala}}, \bibinfo
  {author} {\bibfnamefont {A.}~\bibnamefont {Babiński}}, \bibinfo {author}
  {\bibfnamefont {K.}~\bibnamefont {Watanabe}}, \bibinfo {author}
  {\bibfnamefont {T.}~\bibnamefont {Taniguchi}}, \bibinfo {author}
  {\bibfnamefont {C.}~\bibnamefont {Faugeras}}, \ and\ \bibinfo {author}
  {\bibfnamefont {M.}~\bibnamefont {Potemski}},\ }\href
  {http://arxiv.org/abs/1902.03962} {\bibfield  {journal} {\bibinfo  {journal}
  {{arXiv:1902.03962} [cond-mat]}\ } (\bibinfo {year} {2019})},\ \bibinfo
  {note} {{arXiv:} 1902.03962}\BibitemShut {NoStop}%
\bibitem [{\citenamefont {Molina-Sánchez}\ \emph {et~al.}(2017)\citenamefont
  {Molina-Sánchez}, \citenamefont {Sangalli}, \citenamefont {Wirtz},\ and\
  \citenamefont {Marini}}]{molina-sanchez_ab_2017}%
  \BibitemOpen
  \bibfield  {author} {\bibinfo {author} {\bibfnamefont {A.}~\bibnamefont
  {Molina-Sánchez}}, \bibinfo {author} {\bibfnamefont {D.}~\bibnamefont
  {Sangalli}}, \bibinfo {author} {\bibfnamefont {L.}~\bibnamefont {Wirtz}}, \
  and\ \bibinfo {author} {\bibfnamefont {A.}~\bibnamefont {Marini}},\ }\href
  {\doibase 10.1021/acs.nanolett.7b00175} {\bibfield  {journal} {\bibinfo
  {journal} {Nano Letters}\ }\textbf {\bibinfo {volume} {17}},\ \bibinfo
  {pages} {4549} (\bibinfo {year} {2017})}\BibitemShut {NoStop}%
\bibitem [{\citenamefont {Selig}\ \emph {et~al.}(2018)\citenamefont {Selig},
  \citenamefont {Berghäuser}, \citenamefont {Richter}, \citenamefont
  {Bratschitsch}, \citenamefont {Knorr},\ and\ \citenamefont
  {Malic}}]{selig_dark_2018}%
  \BibitemOpen
  \bibfield  {author} {\bibinfo {author} {\bibfnamefont {M.}~\bibnamefont
  {Selig}}, \bibinfo {author} {\bibfnamefont {G.}~\bibnamefont {Berghäuser}},
  \bibinfo {author} {\bibfnamefont {M.}~\bibnamefont {Richter}}, \bibinfo
  {author} {\bibfnamefont {R.}~\bibnamefont {Bratschitsch}}, \bibinfo {author}
  {\bibfnamefont {A.}~\bibnamefont {Knorr}}, \ and\ \bibinfo {author}
  {\bibfnamefont {E.}~\bibnamefont {Malic}},\ }\href {\doibase
  10.1088/2053-1583/aabea3} {\bibfield  {journal} {\bibinfo  {journal} {{2D}
  Materials}\ }\textbf {\bibinfo {volume} {5}},\ \bibinfo {pages} {035017}
  (\bibinfo {year} {2018})}\BibitemShut {NoStop}%
\bibitem [{\citenamefont {Merkl}\ \emph {et~al.}(2019)\citenamefont {Merkl},
  \citenamefont {Mooshammer}, \citenamefont {Steinleitner}, \citenamefont
  {Girnghuber}, \citenamefont {Lin}, \citenamefont {Nagler}, \citenamefont
  {Holler}, \citenamefont {Schüller}, \citenamefont {Lupton}, \citenamefont
  {Korn}, \citenamefont {Ovesen}, \citenamefont {Brem}, \citenamefont {Malic},\
  and\ \citenamefont {Huber}}]{merkl_ultrafast_2019}%
  \BibitemOpen
  \bibfield  {author} {\bibinfo {author} {\bibfnamefont {P.}~\bibnamefont
  {Merkl}}, \bibinfo {author} {\bibfnamefont {F.}~\bibnamefont {Mooshammer}},
  \bibinfo {author} {\bibfnamefont {P.}~\bibnamefont {Steinleitner}}, \bibinfo
  {author} {\bibfnamefont {A.}~\bibnamefont {Girnghuber}}, \bibinfo {author}
  {\bibfnamefont {K.-Q.}\ \bibnamefont {Lin}}, \bibinfo {author} {\bibfnamefont
  {P.}~\bibnamefont {Nagler}}, \bibinfo {author} {\bibfnamefont
  {J.}~\bibnamefont {Holler}}, \bibinfo {author} {\bibfnamefont
  {C.}~\bibnamefont {Schüller}}, \bibinfo {author} {\bibfnamefont {J.~M.}\
  \bibnamefont {Lupton}}, \bibinfo {author} {\bibfnamefont {T.}~\bibnamefont
  {Korn}}, \bibinfo {author} {\bibfnamefont {S.}~\bibnamefont {Ovesen}},
  \bibinfo {author} {\bibfnamefont {S.}~\bibnamefont {Brem}}, \bibinfo {author}
  {\bibfnamefont {E.}~\bibnamefont {Malic}}, \ and\ \bibinfo {author}
  {\bibfnamefont {R.}~\bibnamefont {Huber}},\ }\href {\doibase
  10.1038/s41563-019-0337-0} {\bibfield  {journal} {\bibinfo  {journal} {Nature
  Materials}\ }\textbf {\bibinfo {volume} {18}},\ \bibinfo {pages} {691}
  (\bibinfo {year} {2019})}\BibitemShut {NoStop}%

\bibitem[{\citenamefont{Semkat et~al.}(2009)\citenamefont{Semkat, Richter,
  Kremp, Manzke, Kraeft, and Henneberger}}]{semkat_ionization_2009}
\bibinfo{author}{\bibfnamefont{D.}~\bibnamefont{Semkat}},
  \bibinfo{author}{\bibfnamefont{F.}~\bibnamefont{Richter}},
  \bibinfo{author}{\bibfnamefont{D.}~\bibnamefont{Kremp}},
  \bibinfo{author}{\bibfnamefont{G.}~\bibnamefont{Manzke}},
  \bibinfo{author}{\bibfnamefont{W.-D.} \bibnamefont{Kraeft}},
  \bibnamefont{and}
  \bibinfo{author}{\bibfnamefont{K.}~\bibnamefont{Henneberger}},
  \bibinfo{journal}{Physical Review B} \textbf{\bibinfo{volume}{80}},
  \bibinfo{pages}{155201} (\bibinfo{year}{2009}),
  \urlprefix\url{http://link.aps.org/doi/10.1103/PhysRevB.80.155201}.

\bibitem[{\citenamefont{Lyddane et~al.}(1941)\citenamefont{Lyddane, Sachs, and
  Teller}}]{lyddane_polar_1941}
\bibinfo{author}{\bibfnamefont{R.~H.} \bibnamefont{Lyddane}},
  \bibinfo{author}{\bibfnamefont{R.~G.} \bibnamefont{Sachs}}, \bibnamefont{and}
  \bibinfo{author}{\bibfnamefont{E.}~\bibnamefont{Teller}},
  \bibinfo{journal}{Physical Review} \textbf{\bibinfo{volume}{59}},
  \bibinfo{pages}{673} (\bibinfo{year}{1941}),
  \urlprefix\url{https://link.aps.org/doi/10.1103/PhysRev.59.673}.


\end{thebibliography}

\section{Appendix}

\subsection{Identification of screened Coulomb interaction and phonon propagator}

We demonstrate the replacement of the screened Coulomb interaction by a phonon propagator given in Eq.~(2) of the manuscript for a well-known limiting case.
The simplest self-energy describing carrier-carrier interaction via a screened Coulomb potential is the GW self-energy that is obtained in random-phase-approximation (RPA)
from Hedin's equations \cite{hedin_new_1965}. In the formalism of nonequilibrium Green functions the retarded GW self-energy is given by \cite{semkat_ionization_2009}
\begin{equation}
 \begin{split}
 & \Sigma_{\bk\lambda}^{\textrm{GW},\textrm{ret}}(\omega) = \Sigma_{\bk\lambda}^{\textrm{Hartree-Fock}} + i\hbar\int_{-\infty}^{\infty}\frac{d\omega'}{2\pi}\\
            \sum_{\bq}&\frac{(1-F^{\lambda}(E^{\lambda}_{\bq})+ n_{\textrm{B}}(\omega'))2i\,V_{\bk-\bq}\,\textrm{Im}\,\varepsilon^{-1}_{\bk-\bq}(\omega')}{\hbar\omega-E^{\lambda}_{\bq}+i\gamma^{\lambda}_{\bq}-\hbar\omega'}
     \,
 \end{split}
\label{eq:GW}
\end{equation}
with quasi-particle energies $ E^{\lambda}_{\bk}$ and dampings $ \gamma^{\lambda}_{\bq}$ for carriers with momentum $\bk$ in band $\lambda$. $F^{\lambda}(\omega)$ are Fermi functions describing the distribution of carriers, while $n_{\textrm{B}}(\omega) $ are Bose functions belonging to the bosonic excitations contained in the loss function $\textrm{Im}\,\varepsilon^{-1}_{\bq}(\omega)$. Inserting the loss function according to 
\begin{equation}
 \begin{split}
V_{\bq}\textrm{Im}\,\varepsilon^{-1}_{\bq}(\omega) =-\pi \left|g_{\bq}\right|^2\left(\delta(\hbar\omega-\hbar\Omega_{\bq}) - \delta(\hbar\omega+\hbar\Omega_{\bq})\right)\,
\label{eq:lossf}
\end{split}
\end{equation}
with matrix elements $g_{\bq}$ and energies $\hbar\Omega_{\bq}$ we obtain for the self-energy beyond Hartree-Fock:
\begin{equation}
 \begin{split}
 \Sigma_{\bk\lambda}^{\textrm{RPA},\textrm{ret}}(\omega) = \sum_{\bq}\Bigg\lbrace&\frac{(1-F^{\lambda}(E^{\lambda}_{\bq})+ n_{\textrm{B}}(\Omega_{\bk-\bq}))|g_{\bk-\bq}|^2}{\hbar\omega-E^{\lambda}_{\bq}+i\gamma^{\lambda}_{\bq}-\Omega_{\bk-\bq}} \\
 +&\frac{(F^{\lambda}(E^{\lambda}_{\bq})+ n_{\textrm{B}}(\Omega_{\bk-\bq}))|g_{\bk-\bq}|^2}{\hbar\omega-E^{\lambda}_{\bq}+i\gamma^{\lambda}_{\bq}+\Omega_{\bk-\bq}}\Bigg\rbrace
 \,
     \,.
 \end{split}
\label{eqphon_RPA}
\end{equation}
Here we have evaluated the Delta distributions of the phonon propagator using the relation $n_{\textrm{B}}(-\omega)=-1-n_{\textrm{B}}(\omega)$. This self-energy describes quasi-particle renormalizations due to carrier-phonon interaction in RPA, see Eq.~(1) in Ref.~\citenum{molina-sanchez_temperature-dependent_2016}. The two terms in the loss function (\ref{eq:lossf}) lead to phonon emission and absorption contributions to the self-energy.

In a second step, we assume the case of an ideal 2d layer that is not embedded into a dielectric environment. The layer itself may host optical phonons so that the dielectric function is given by \cite{haug_electron_1984}
\begin{equation}
 \begin{split}
 \varepsilon(\omega)=\varepsilon_{\infty}+\frac{s^2}{\omega_0^2-\omega^2-i\gamma\omega}\,
 \end{split}
\label{eq:2d_DF_TO}
\end{equation}
with $\omega_0=\omega_{\textrm{TO}}$ the transversal-optical phonon energy and the coupling strength $s^2=(\varepsilon_{\textrm{static}}-\varepsilon_{\infty})\omega_{\textrm{TO}}^2$ as obtained from the static and high-frequency limits of $\varepsilon(\omega)$. Calculating the inverse of $\varepsilon(\omega)$ and using the 2d Coulomb potential $V_{\bq}=\frac{e^2}{2\varepsilon_0 q}$, we obtain from Eq.~(2):
\begin{equation}
 \begin{split}
 |g_{\bq}|^2&=\frac{e^2}{2\varepsilon_0 q}\frac{(\varepsilon_{\textrm{static}}-\varepsilon_{\infty})\omega_{\textrm{TO}}^2}{\varepsilon_{\infty}^2}\frac{\hbar}{2\Omega_{\bq}}\,,\\
 \Omega_{\bq}&=\sqrt{\omega_{\textrm{TO}}^2+\frac{(\varepsilon_{\textrm{static}}-\varepsilon_{\infty})\omega_{\textrm{TO}}^2}{\varepsilon_{\infty}}}\\
	  &=\omega_{\textrm{TO}}\sqrt{\frac{\varepsilon_{\textrm{static}}}{\varepsilon_{\infty}}}=\omega_{\textrm{LO}}\,.
 \end{split}
\label{eq:g2}
\end{equation}
In the last line, we used the Lyddane-Sachs-Teller relation \cite{lyddane_polar_1941}. For the matrix elements we thus obtain
\begin{equation}
 \begin{split}
 |g_{\bq}|^2=\frac{e^2}{2\varepsilon_0 q}\left(\frac{1}{\varepsilon_{\infty}}-\frac{1}{\varepsilon_{\textrm{static}}}\right)\frac{\hbar\omega_{\textrm{LO}}}{2}\,,
 \end{split}
\label{eq:g2_final}
\end{equation}
which corresponds to the standard Fröhlich model for carrier-LO-phonon coupling.

\subsection{Dielectric function of hBN}

\begin{figure}[h!t]
\centering
\includegraphics[width=\columnwidth]{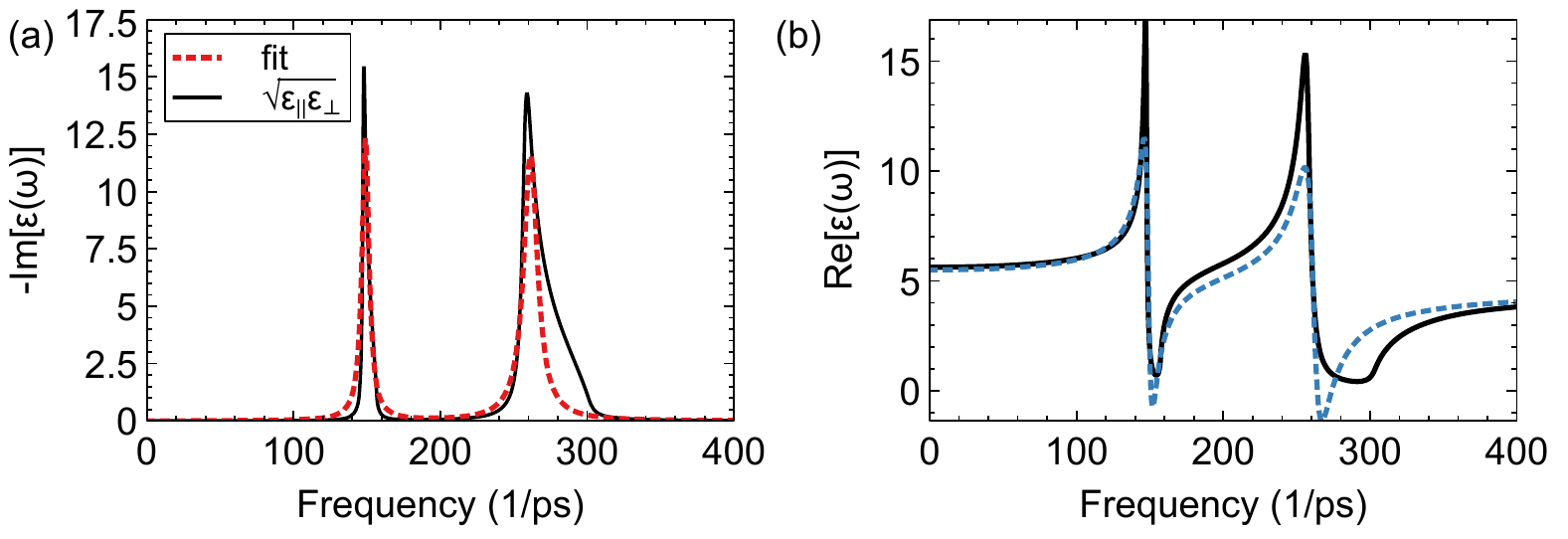}
\caption{Fit to the effective dielectric function of hBN $\varepsilon_{\textrm{eff}}=\sqrt{\varepsilon_{\parallel}\varepsilon_{\perp}}$ using the double-oscillator model given in Eq.~(\ref{eq:double_osc}).}
\label{fig:hBN_fit}
\end{figure}

According to \cite{mele_screening_2001}, anisotropic dielectric screening can be effectively described by the dielectric function $\varepsilon_{\textrm{eff}}=\sqrt{\varepsilon_{\parallel}\varepsilon_{\perp}}$, where the full dielectric tensor has the form $\varepsilon_{ij}=\varepsilon_{\parallel}\delta_{xx}+\varepsilon_{\parallel}\delta_{yy}+\varepsilon_{\perp}\delta_{zz}$. The dielectric function of hBN including one in-plane and out-of-plane optical phonon, respectively, is given in Ref.~\citenum{geick_normal_1966}. Due to the geometric mean, the effective dielectric function obtains an asymmetric line shape, see Fig.~\ref{fig:hBN_fit}. However, the dielectric function can be appropriately reproduced by a model including two harmonic oscillators:
\begin{equation}
\begin{split}
\varepsilon^{\textrm{hBN}}(\omega)=\varepsilon_{\infty}+\frac{s_1^2}{\omega_{0,1}^2-\omega^2-i\gamma_1\omega}+\frac{s_2^2}{\omega_{0,2}^2-\omega^2-i\gamma_2\omega}\,.
\label{eq:double_osc}
\end{split}
\end{equation}
A least-square fit yields the parameters $\varepsilon_{\infty}=4.5$, $s_1=68$ meV, $s_2=123$ meV, $\omega_{0,1}=98$ meV, $\omega_{0,2}=172$ meV, $\gamma_1=3.8$ meV and $\gamma_2=7.6$ meV.

\subsection{General expression for carrier-phonon coupling in a dielectric environment}

\begin{figure}[h!t]
\centering
\includegraphics[width=0.5\columnwidth]{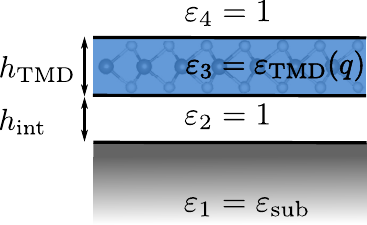}
\caption{Schematic of the asymmetric dielectric heterostructure: A two-dimensional layer is placed on a substrate with an inter-layer gap in between.}
\label{fig:struct}
\end{figure}

Expression (11) in the main text for carrier-phonon coupling matrix elements $g_{\bq}$ and bosonic resonance energies $\hbar\Omega_{\bq}$ can be transferred to an asymmetric dielectric environment of the two-dimensional layer, which is shown in Fig.~\ref{fig:struct}. To this end, we rely on the heterostructure dielectric function given as Eq.~(3) in Ref.~\citenum{florian_dielectric_2018}.
%
Using the definition $\tilde{\varepsilon}_i=\frac{\varepsilon_{i+1}-\varepsilon_i}{\varepsilon_{i+1}+\varepsilon_i}$, Eq.~(11) can be reused, with $K=bc-ad$ and redefined parameters $a$, $b$, $c$ and $d$:
\begin{equation}
\begin{split}
a &= 1+\tilde{\varepsilon}_2\beta+\tilde{\varepsilon}_3\alpha^2\beta+\tilde{\varepsilon}_2\tilde{\varepsilon}_3\alpha^2\\
b &= 1-\tilde{\varepsilon}_2\beta-\tilde{\varepsilon}_3\alpha^2\beta+\tilde{\varepsilon}_2\tilde{\varepsilon}_3\alpha^2 \\
c &= 1+\alpha\beta+(\tilde{\varepsilon}_2-\tilde{\varepsilon}_3)\alpha+\tilde{\varepsilon}_2\beta-\tilde{\varepsilon}_3\alpha^2\beta-\tilde{\varepsilon}_2\tilde{\varepsilon}_3(\alpha^2+\alpha\beta) \\
d &= 1-\alpha\beta+(\tilde{\varepsilon}_2-\tilde{\varepsilon}_3)\alpha-\tilde{\varepsilon}_2\beta+\tilde{\varepsilon}_3\alpha^2\beta-\tilde{\varepsilon}_2\tilde{\varepsilon}_3(\alpha^2-\alpha\beta) \,.
\label{eq:abcd}
\end{split}
\end{equation}
%

%

\subsection{Temperature dependence of exciton-substrate-phonon coupling}
\begin{figure}[h!t]
\centering
\includegraphics[width=\columnwidth]{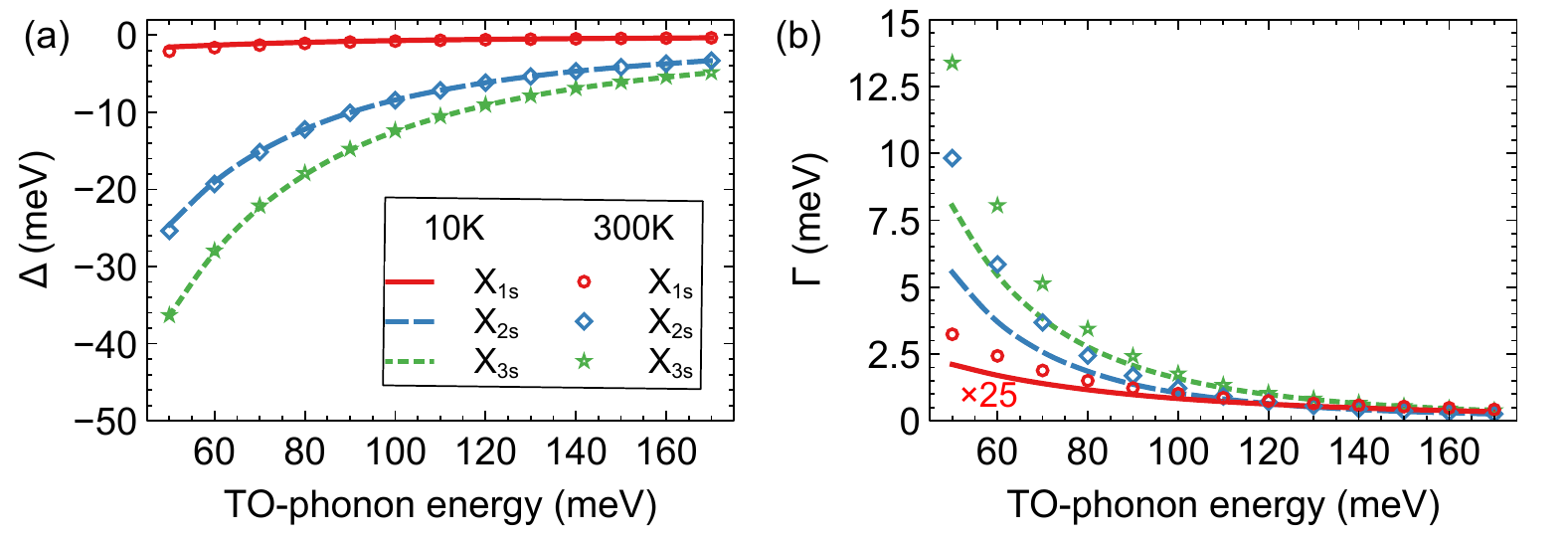}
\caption{Temperature dependence of exciton energy renormalizations \textbf{(a)} and exciton line broadenings \textbf{(b)} induced by carrier-substrate-phonon interaction.}
\label{fig:temp}
\end{figure}
We study the temperature dependence of renormalization effects induced by the interaction of 2d-carriers with optical phonons in the environment by comparing results for $T=300$ K and $T=10$ K, see Fig.~\ref{fig:temp}.
We find that while the linewidth broadening shows some effect especially at low phonon energies, exciton energy renormalizations are practically not affected by temperature. This behavior can be understood by an approximate analytic treatment of the effective exciton-exciton Hamiltonian, see Eq.~(8) in the main text. Assuming a simple $1/\sqrt{q}$-dependence of the matrix elements $g_{\bq}$, a constant resonance frequency $\Omega_{\bq}=\omega_0$ with Bose functions $n_{\bq}=n_0=n(\omega_0)$, and exciton dispersion $E_{\alpha,\bq}=E_{\alpha,\boldsymbol{0}} + \beta q^2 $, the Hamiltonian can be simplified to
\begin{equation}
 \begin{split}
 \big< \nu,\boldsymbol{0} \big| H^{\textrm{eff}} \big| \nu',\boldsymbol{0} \big> & \approx \sum_{\alpha\bq}\frac{\gamma}{q} \\
 &\times\Big(\frac{1+n_0}{E_{\nu',\boldsymbol{0}}-E_{\alpha,\boldsymbol{0}}-\beta q^2-\hbar\omega_0+i\Gamma } \\ &+\frac{n_0}{E_{\nu',\boldsymbol{0}}-E_{\alpha,\boldsymbol{0}}-\beta q^2+\hbar\omega_0+i\Gamma}\Big)
 \\
 &=\sum_{\alpha}\int_0^{\infty}dq q \frac{\gamma}{2\pi q} \\
 &\times\Big(\frac{1+n_0}{E_{\nu',\boldsymbol{0}}-E_{\alpha,\boldsymbol{0}}-\beta q^2-\hbar\omega_0+i\Gamma } \\ &+\frac{n_0}{E_{\nu',\boldsymbol{0}}-E_{\alpha,\boldsymbol{0}}-\beta q^2+\hbar\omega_0+i\Gamma}\Big)\\
 &=\sum_{\alpha}\frac{\gamma}{2\pi}(-\frac{\pi}{2}) \\
 &\times\Big(\frac{1+n_0}{\sqrt{\beta(\hbar\omega_0+E_{\alpha,\boldsymbol{0}}- E_{\nu',\boldsymbol{0}}-i\Gamma )}} \\ &+\frac{n_0}{\sqrt{-\beta(\hbar\omega_0+E_{\nu',\boldsymbol{0}}- E_{\alpha,\boldsymbol{0}}+i\Gamma )}}\Big)\,.
\end{split}
\label{eq:Heff_approx}
\end{equation}
We focus on scattering processes within the same exciton branch, neglecting inter-exciton transitions with $\alpha\neq\nu$. Taking the limit $\Gamma\ll\omega_0$, we obtain for the diagonal matrix elements
\begin{equation}
 \begin{split}
 \Delta E_{\nu,\boldsymbol{0}}&=\big< \nu,\boldsymbol{0} \big| H^{\textrm{eff}} \big| \nu,\boldsymbol{0} \big> \\
 &\approx-\frac{\gamma}{4}\Big(\frac{1+n_0}{\sqrt{\beta\hbar\omega_0}} +i\frac{n_0}{\sqrt{\beta\hbar\omega_0}}\Big)\,.
\end{split}
\label{eq:Delta_E_approx}
\end{equation}
The real part describes a red shift of exciton energies, while the negative imaginary part accounts for linewidth broadening. One can see that the broadening is approximately proportional to the phonon population, which makes it very sensitive to the lattice temperature, while energy shifts obtain a contribution due to phonon emision at any temperature. Moreover, a scaling of renormalizations with the inverse square root of the substrate phonon energy is found. It is expected that a more realistic momentum dependence of the exciton-phonon matrix elements $g_{\bq}$ yields modifications of the simple $1/\sqrt{\omega_0}$ behavior. However, the general inverse power law scaling found from the full calculation can be roughly understood from the simplified calculation.

\end{document}